\documentclass[english,aps,prb,reprint]{revtex4-2}
\usepackage[T1]{fontenc}
\usepackage[latin9]{inputenc}
\usepackage{bm}
\usepackage{amsmath}
\usepackage{graphicx}

\makeatletter
\usepackage[all]{xy}
\usepackage{mathtools}
\usepackage{variables}
\usepackage[bookmarks=false]{hyperref}

\makeatother

\usepackage{babel}
\begin{document}
\title{A gauge constrained algorithm of VDAT at $\discn=3$ for the multi-orbital
Hubbard model}
\author{Zhengqian Cheng and Chris A. Marianetti}
\affiliation{Department of Applied Physics and Applied Mathematics, Columbia University,
New York, NY 10027}
\date{\today}
\begin{abstract}
The recently developed variational discrete action theory (VDAT) provides
a systematic variational approach to the ground state of the quantum
many-body problem, where the quality of the solution is controlled
by an integer $\discn$, and increasing $\discn$ monotonically approaches
the exact solution. VDAT can be exactly evaluated in the $d=\infty$
multi-orbital Hubbard model using the self-consistent canonical discrete
action theory (SCDA), which requires a self-consistency condition
for the integer time Green's functions. Previous work demonstrates
that $\discn=3$ accurately captures multi-orbital Mott/Hund physics
at a cost similar to the Gutzwiller approximation. Here we employ
a gauge constraint to automatically satisfy the self-consistency condition
of the SCDA at $\discn=3$, yielding an even more efficient algorithm
with enhanced numerical stability. We derive closed form expressions
of the gauge constrained algorithm for the multi-orbital Hubbard model
with general density-density interactions, allowing VDAT at $\discn=3$
to be straightforwardly applied to the seven orbital Hubbard model.
We present results and a performance analysis using $\discn=2$ and
$\discn=3$ for the $\textrm{SU}(2\textrm{N}_{\textrm{orb}})$ Hubbard
model in $d=\infty$ with $\textrm{N}_{\textrm{orb}}=2-8$, and compare
to numerically exact dynamical mean-field theory solutions where available.
The developments in this work will greatly facilitate the application
of VDAT at $\discn=3$ to strongly correlated electron materials.
\end{abstract}
\maketitle

\section{Introduction}

The recently developed variational discrete action theory (VDAT) \cite{Cheng2021195138,Cheng2021206402}
has emerged as a powerful tool to study the ground state of the multi-orbital
Hubbard model \cite{Cheng2022205129}, which can be considered as
a minimal model for a wide class of strongly correlated electron materials
\cite{Imada19981039,Kotliar2006865}. VDAT consists of two central
components: the sequential product density matrix (SPD) ansatz and
the discrete action theory to evaluate observables under the SPD.
The accuracy of the SPD is controlled by an integer $\discn$, and
the SPD monotonically approaches the exact solution for increasing
$\discn$. In the context of the Hubbard model, the SPD recovers most
well known variational wavefunctions \cite{Cheng2021195138}: $\discn=1$
recovers the Hartree-Fock wave function, $\discn=2$ recovers the
Gutzwiller wave function \cite{Gutzwiller1963159,Gutzwiller1964923,Gutzwiller19651726},
and $\discn=3$ recovers the Gutzwiller-Baeriswyl \cite{Otsuka19921645}
and Baeriswyl-Gutzwiller wavefunctions \cite{Dzierzawa19951993}.
The discrete action theory can be viewed as an integer time generalization
of the imaginary time path integral, yielding an integer time generalization
of the Green's function and Dyson equation \cite{Cheng2021195138}.
For $d=\infty$, the SPD can be exactly evaluated using the self-consistent
canonical discrete action (SCDA) \cite{Cheng2021195138,Cheng2021206402}.
VDAT within the SCDA offers a paradigm shift away from the dynamical
mean-field theory (DMFT) \cite{Georges199613,Kotliar2006865,Vollhardt20121},
allowing the exact solution of the ground state properties of the
$d=\infty$ Hubbard model to be systematically approached within the
wave function paradigm. The computational cost of VDAT grows with
$\discn$, at an exponential scaling for an exact evaluation and a
polynomial scaling for a numerical evaluation using Monte-Carlo, so
rapid convergence with $\discn$ is important if VDAT is to be a practical
alternative to DMFT. VDAT using $\discn=2,3,4$ has been applied to
the single orbital Anderson impurity model on a ring \cite{Cheng2021206402},
the $d=\infty$ single orbital Hubbard model \cite{Cheng2021206402},
and the $d=\infty$ two orbital Hubbard model \cite{Cheng2022205129},
and in all cases $\discn=3$ yields accurate results as compared to
the numerically exact solutions. This success is particularly nontrivial
in the two orbital problem, where complex local interactions including
the Hubbard $U$, Hund $J$, and crystal field $\Delta$ were studied
over all parameter space. Therefore, VDAT within the SCDA at $\discn=3$
provides a minimal and accurate description of the two orbital Hubbard
model, but with a computational cost that is comparable to $\discn=2$,
which recovers the Gutzwiller approximation \cite{Gutzwiller19651726}
(GA) and slave boson mean-field theories \cite{Kotliar19861362,Bunemann2007193104,Piefke2018125154}.
The fact that VDAT within the SCDA at $\discn=3$ resolves all the
limitations of the Gutzwiller approximation and the slave boson mean-field
theories without substantially increasing the computational cost motivates
a deeper understanding of how the SCDA works.

The SCDA provides a route for exactly evaluating the SPD in $d=\infty$,
and the SCDA can be viewed as the integer time analogue of DMFT \cite{Cheng2021195138,Cheng2022205129}.
While DMFT maps the Hubbard model to a self-consistently determined
Anderson impurity model, the SCDA maps the SPD to a self-consistently
determined canonical discrete action (CDA), parametrized by the corresponding
non-interacting integer time Green's function $\mathcal{G}$, which
implicitly depends on the variational parameters of the SPD. While
the DMFT self-consistency condition only needs to be executed once,
the SCDA self-consistency condition must be executed for every choice
of variational parameters during the minimization. Previously, we
proposed an approach to mitigate this issue by simultaneously minimizing
the variational parameters and updating $\mathcal{G}$, and demonstrated
it to be efficient for the two band Hubbard model \cite{Cheng2022205129}.
However, for some regions of parameter space, such as in the large
polarization regime, a very small step size is needed to maintain
numerical stability. Such problems are partially due to inaccuracies
of $\mathcal{G}$ within the iteration process, given that $\mathcal{G}$
is only highly precise when the fixed point is reached. Therefore,
it would be advantageous if the self-consistency condition of the
SCDA could be automatically satisfied.

Given that $\mathcal{N}=2$ recovers the GA, it is interesting to
recall how the GA automatically satisfies the SCDA self-consistency
condition. As previously demonstrated \cite{Cheng2021195138}, the
GA has a prescribed form of $\mathcal{G}$ which is fully determined
by imposing that the non-interacting and interacting local single
particle density matrices are identical, which we refer to as the
Gutzwiller gauge. While the SCDA at $\mathcal{N}=2$ can evaluate
an SPD with arbitrary variational parameters, the GA is only valid
when the SPD satisfies the Gutzwiller gauge. Therefore, the GA is
a special case of the SCDA at $\mathcal{N}=2$, but the Gutzwiller
gauge does not limit the variational power of the SPD due to the gauge
freedom of the SPD \cite{Cheng2022205129}. In summary, the GA provides
an important lesson for numerically simplifying the SCDA at $\mathcal{N}=2$
by exploiting the gauge freedom of the SPD, converting the problem
of solving for $\mathcal{G}$ into a constraint on the variational
parameters of the SPD. In this paper, we demonstrate that the lessons
of the GA can be generalized to $\discn=3$, which quantitatively
captures the Mott and Hund physics of the multi-orbital Hubbard model
\cite{Cheng2022205129}. 

In order to demonstrate the power of the gauge constrained implementation
of the SCDA at $\mathcal{N}=3$, we study the $\textrm{SU}(2\textrm{N}_{\textrm{orb}})$
Hubbard model in $d=\infty$ with $\textrm{N}_{\textrm{orb}}=2-8$.
Our successful execution of these calculations demonstrates the viability
of applying VDAT at $\mathcal{N}=3$ to crystals bearing $d$ or $f$
electrons. Moreover, the $\textrm{SU}(2\textrm{N}_{\textrm{orb}})$
Hubbard model is interesting in its own right, given that experiments
on ultracold atoms in an optical lattice can realize the $\textrm{SU}(2\textrm{N}_{\textrm{orb}})$
Hubbard model\cite{Taie2012825,He2020012028,Ozawa2018225303,Sonderhouse20201038,Schafer2020411,Zhang2020213}.
Therefore, VDAT should serve as a reliable tool for understanding
and interpreting such experimental measurements. 

The structure of this paper is as follows. Sec. \ref{sec:Gauge-constrained-SCDA}
presents the gauge constrained algorithm of the SCDA at $\discn=3$,
with Sec. \ref{subsec:Gauge_overview} providing a high level overview
of the entire algorithm, including all key equations, while the remaining
subsections provide detailed derivations. Sec. \ref{sec:Results-for-SU(2N)}
presents results for the $\textrm{SU}(2\textrm{N}_{\textrm{orb}})$
Hubbard model in $d=\infty$ with $\textrm{N}_{\textrm{orb}}=2-8$.

\section{Gauge constrained algorithm at $\discn$=3\label{sec:Gauge-constrained-SCDA}}

\subsection{Overview\label{subsec:Gauge_overview}}

The goal of this subsection is to provide an overview of the gauge
constrained algorithm for the SCDA at $\discn=3$, and Secs. \ref{subsec:A-block},
\ref{subsec:A-block_momentum_part}, and \ref{subsec:BCD-block} will
derive all details of the procedure. We begin by highlighting how
the SCDA exactly evaluates the SPD in $d=\infty$ \cite{Cheng2021195138,Cheng2021206402,Cheng2022205129}.
Consider a fermionic lattice model having a Hamiltonian
\begin{equation}
\hat{H}=\hat{K}+\hat{H}_{loc}=\sum_{k\alpha\sigma}\epsilon_{k\alpha\sigma}\hat{n}_{k\alpha\sigma}+\sum_{i}\hat{H}_{loc;i},
\end{equation}
where $i=1,\dots,\textrm{N}_{\textrm{site}}$ enumerates over the
lattice sites, $k=1,\dots,\textrm{N}_{\textrm{site}}$ enumerates
over the $k$-points, $\alpha=1,\dots,\textrm{N}_{\textrm{orb}}$
enumerates over the orbitals, and $\sigma$ enumerates over spin.
The G-type SPD for $\discn=3$ can be motivated from the following
variational wave function 
\begin{align}
 & \exp\left(\sum_{k\alpha\sigma}\gamma_{k\alpha\sigma}\hat{n}_{k\alpha\sigma}\right)\prod_{i}\hat{P}_{i}\left(u\right)|\Psi_{0}\rangle,
\end{align}
where $\{\gamma_{k\alpha\sigma}\}$ is the set of non-interacting
variational parameters, $u=\{u_{i\Gamma}\}$ is the set of interacting
variational parameters, $\hat{P}_{i}\left(u\right)=\sum_{\Gamma}u_{i\Gamma}\hat{P}_{i\Gamma}$,
and $|\Psi_{0}\rangle$ is a non-interacting variational wavefunction;
and both $\{\gamma_{k\alpha\sigma}\}$ and $\{u_{i\Gamma}\}$ are
real numbers. The index $\Gamma$ enumerates a set of many-body operators
$\{\hat{P}_{i\Gamma}\}$ local to site $i$. The $\{\hat{P}_{i\Gamma}\}$
used for an $\hat{H}_{loc;i}$ containing only density-density type
interactions is given in Eq. \ref{eq:SUN_projector}, while the general
case is given in Eq. \ref{eq:general_projector}. It will be important
to rewrite the above wave function as a density matrix, yielding the
G-type SPD \cite{Cheng2021195138} for $\discn=3$
\begin{align}
\spd & =\hat{\mathcal{P}}_{1}\hat{\mathcal{P}}_{2}\hat{\mathcal{P}}_{3}=\left(\hat{K}_{1}\hat{P}_{1}\right)\left(\hat{K}_{2}\hat{P}_{1}\right)\left(\hat{K}_{1}\right),\label{eq:SPD}
\end{align}
where $\hat{K}_{1}=\exp\left(\sum_{k\alpha\sigma}\gamma_{k\alpha\sigma}\hat{n}_{k\alpha\sigma}\right)$,
$\hat{P}_{1}=\prod_{i}\hat{P}_{i}\left(u\right)$, and $\hat{K}_{2}=|\Psi_{0}\rangle\langle\Psi_{0}|$.
Here we have chosen $\hat{K}_{1}$ to be diagonal in $k\alpha\sigma$,
while the most general case is addressed in Ref. \cite{Cheng2022205129}. 

Evaluating expectation values under the SPD is highly nontrivial,
and we have developed the discrete action theory \cite{Cheng2021195138}
to formalize the problem in a manner which is amenable to systematic
approximations. A key idea of the discrete action theory is the equivalence
relation between an integer time correlation function and a corresponding
expectation value in the compound space. An operator operator $\hat{O}$
in the original space is promoted to the compound space with a given
integer time index $\tau$, denoted as $\barhat[O]^{(\tau)}$ \cite{Cheng2021195138}.
For the total energy with $\discn=3$, this equivalence is given as 

\begin{align}
\langle\hat{H}\rangle_{\spd} & =\langle\barhat[H]^{\left(\discn\right)}\rangle_{\spdcs},
\end{align}
where $\spdcs=\spdcs_{0}\prod_{i}\barhat[P]_{i}$ is the discrete
action of the SPD, $\spdcs_{0}=\barhat[Q]\barhat[K]_{1}^{\left(1\right)}\barhat[K]_{2}^{\left(2\right)}\barhat[K]_{1}^{\left(3\right)}$
is the non-interacting discrete action, $\barhat[Q]$ is the integer
time translation operator \cite{Cheng2021195138,Cheng2022205129},
and the interacting projector for site $i$ is
\begin{equation}
\barhat[P]_{i}=\barhat[P]_{i}^{\left(1\right)}(u)\barhat[P]_{i}^{\left(2\right)}(u)=\sum_{\Gamma\Gamma'}u_{i\Gamma}u_{i\Gamma'}\barhat[P]_{i\Gamma}^{\left(1\right)}\barhat[P]_{i\Gamma'}^{\left(2\right)}.\label{eq:local_int_proj}
\end{equation}
 An important previous result is that the expectation value in the
compound space can be exactly evaluated for $d=\infty$ using the
self-consistent canonical discrete action theory (SCDA) \cite{Cheng2021206402,Cheng2021195138}. 

Given the common scenario of translation symmetry, the SCDA can be
presented in terms of two auxiliary effective discrete actions parameterized
by $2\textrm{N}_{\text{orb}}\discn\times2\textrm{N}_{\text{orb}}\discn$
matrices $\bm{S}_{loc}$ and $\bm{\mathcal{G}}$, given as 
\begin{align}
 & \langle\barhat[H]^{\left(\discn\right)}\rangle_{\spdcs}=\langle\barhat[K]^{\left(\discn\right)}\rangle_{\barhat[\rho]_{K}}+\textrm{N}_{\textrm{site}}\langle\barhat[H]_{loc;i}^{\left(\discn\right)}\rangle_{\barhat[\rho]_{loc;i}},\label{eq:h_energy}\\
 & \barhat[\rho]_{K}=\spdcs_{0}\exp\Big(-\sum_{i}\ln\bm{S}_{loc}^{T}\cdot\barhat[\bm{n}]_{i}\Big),\label{eq:rhoK}\\
 & \barhat[\rho]_{loc;i}=\exp\Big(-\ln\left(\bm{\mathcal{G}}^{-1}-\boldsymbol{1}\right)^{T}\cdot\barhat[\bm{n}]_{i}\Big)\barhat[P]_{i},\label{eq:rholoc}
\end{align}
where $[\barhat[\bm{n}]_{i}]_{\alpha\sigma\tau,\alpha'\sigma'\tau'}=\barhat[a]_{i\alpha\sigma}^{\dagger(\tau)}\barhat[a]_{i\alpha'\sigma'}^{(\tau')}$,
the dot product operation is defined as $\boldsymbol{a}\cdot\hat{\boldsymbol{b}}\equiv\sum_{\ell\ell'}[\boldsymbol{a}]_{\ell\ell'}[\hat{\boldsymbol{b}}]_{\ell\ell'}$,
the discrete action $\barhat[\rho]_{K}$ is used to compute all single-particle
integer time Green's functions, and $\barhat[\rho]_{loc;i}$ is used
to compute all N-particle integer time Green's functions local to
site $i$. The $\bm{\mathcal{G}}$ and $\bm{S}_{loc}$ must satisfy
the following two self-consistency conditions
\begin{align}
 & \left(\boldsymbol{g}_{loc}^{-1}-\boldsymbol{1}\right)=\left(\bm{\mathcal{G}}^{-1}-\boldsymbol{1}\right)\boldsymbol{S}_{loc},\label{eq:dysonlocal}\\
 & \bm{g}_{loc}=\bm{g}'_{loc},\label{eq:scda_delta}
\end{align}
where $\bm{g}_{loc}=\left\langle \barhat[\bm{n}]_{i}\right\rangle _{\barhat[\rho]_{loc;i}}$
and $\bm{g}'_{loc}=\left\langle \barhat[\bm{n}]_{i}\right\rangle _{\barhat[\rho]_{K}}$. 

One challenge posed by the SCDA is that the self-consistency condition
must be satisfied for a given choice of variational parameters, which
makes the minimization over the variational parameters nontrivial.
An efficient algorithm for VDAT within the SCDA was proposed for the
multi-orbital Hubbard model for general $\discn$, referred to as
the decoupled minimization algorithm, and implemented in the two orbital
Hubbard model up to $\discn=4$ \cite{Cheng2022205129}. The decoupled
minimization algorithm begins with an initial choice of variational
parameters $\{\gamma_{k\alpha\sigma}\},u$ and an initial choice for
$\bm{\mathcal{G}}$, which determines the discrete action $\barhat[\rho]_{loc;i}$
(Eq. \ref{eq:rholoc}) which yields $\boldsymbol{g}_{loc}$. Using
the discrete Dyson equation (Eq. \ref{eq:dysonlocal}), $\boldsymbol{S}_{loc}$
can be computed from $\bm{\mathcal{G}}$ and $\boldsymbol{g}_{loc}$.
Then the discrete action $\barhat[\rho]_{K}$ (Eq. \ref{eq:h_energy})
can be used to compute $\boldsymbol{g}_{loc}'$. Using $\boldsymbol{g}_{loc}=\boldsymbol{g}_{loc}'$
in the discrete Dyson equation, a new $\bm{\mathcal{G}}$ can be obtained.
During this self-consistency cycle, relevant first order derivatives
with regard to $\{\gamma_{k\alpha\sigma}\}$, $u$, and $\bm{\mathcal{G}}$
can be computed and two effective models can be constructed to update
$\{\gamma_{k\alpha\sigma}\}$ and $u$. This entire procedure is iterated
until $\{\gamma_{k\alpha\sigma}\}$, $u$, and $\bm{\mathcal{G}}$
are self-consistent. In a given iteration before reaching self-consistency,
the energy and its gradients contain errors due to a deviation from
the SCDA self-consistency condition, which can yield slow convergence
in some regions of parameter space. Automatically satisfying the SCDA
would yield a dramatic advantage when minimizing over the variational
parameters. 

In previous work \cite{Cheng2021195138}, we demonstrated that the
gauge freedom of the SPD can be used to automatically satisfy the
SCDA self-consistency condition at $\discn=2$, which recovers the
Gutzwiller approximation, and here we extend this line of reasoning
to $\discn=3$. For simplicity, we use a restricted form of the SPD,
where the kinetic projector is diagonal in k-space and $\hat{P}_{i}\left(u\right)$
does not introduce off-diagonal terms at the level of the single particle
density matrix. Therefore, $\bm{\mathcal{G}}$, $\boldsymbol{S}_{loc}$,
and $\boldsymbol{g}_{loc}$ all have the form $[\boldsymbol{g}_{loc}]_{\alpha\sigma\tau,\alpha'\sigma'\tau'}=\delta_{\alpha\alpha'}\delta_{\sigma\sigma'}[\boldsymbol{g}_{loc}]_{\alpha\sigma\tau,\alpha\sigma\tau'}$,
and the integer time Green's functions of each spin orbital are described
by a $3\times3$ matrix. We begin by partitioning a local integer
time $3\times3$ matrix $\bm{M}_{\alpha\sigma}$ for a given spin
orbital into submatrices as:

\begin{align}
\boldsymbol{M}_{\alpha\sigma} & =\left(\begin{array}{cc|c}
[\boldsymbol{M}_{\alpha\sigma}]_{11} & [\boldsymbol{M}_{\alpha\sigma}]_{12} & [\boldsymbol{M}_{\alpha\sigma}]_{13}\\{}
[\boldsymbol{M}_{\alpha\sigma}]_{21} & [\boldsymbol{M}_{\alpha\sigma}]_{22} & [\boldsymbol{M}_{\alpha\sigma}]_{23}\\
\hline \,[\boldsymbol{M}_{\alpha\sigma}]_{31} & [\boldsymbol{M}_{\alpha\sigma}]_{32} & [\boldsymbol{M}_{\alpha\sigma}]_{33}
\end{array}\right)\\
 & =\begin{pmatrix}\boldsymbol{M}_{\alpha\sigma;A} & \boldsymbol{M}_{\alpha\sigma;B}\\
\boldsymbol{M}_{\alpha\sigma;C} & \boldsymbol{M}_{\alpha\sigma;D}
\end{pmatrix},\label{eq:block_matrix}
\end{align}
where $\bm{M}$ can be $\boldsymbol{\mathcal{G}}$, $\boldsymbol{S}_{loc}$,
$\boldsymbol{g}_{loc}$, and $\boldsymbol{g}_{loc}'$. The main idea
is to satisfy the self-consistency condition $\boldsymbol{g}_{loc}=\boldsymbol{g}_{loc}'$
in two stages: first for the A block and then for the B, C, and D
blocks. 

We proceed by outlining the logic and key equations of the first stage,
which is treated in detail in Section \ref{subsec:A-block} and \ref{subsec:A-block_momentum_part}.
The first stage begins by considering $\bm{\mathcal{G}}_{\alpha\sigma;A}$,
a $2\times2$ matrix, which can be parametrized in terms of the single
variable $\mathcal{G}_{\alpha\sigma;12}$ using the gauge freedom
of the SPD, and $\mathcal{G}_{\alpha\sigma;12}$ should now be regarded
as an independent variational parameter. The $\bm{g}_{loc;A}$ and
$\bm{S}_{loc}$ can be determined as a function of the sets $\mathcal{G}_{12}=\left\{ \mathcal{G}_{\alpha\sigma;12}\right\} $
and $u=\{u_{i\Gamma}\}$, though we suppress the function arguments
$\mathcal{G}_{12}$ and $u$ for brevity. The local density is also
a function of $\mathcal{G}_{12}$ and $u$, defined as $n_{\alpha\sigma}(\mathcal{G}_{12},u)=\left[g_{loc;\alpha\sigma}\right]_{22}$.
For a given $\alpha\sigma$, the $\bm{S}_{loc;\alpha\sigma}$ can
be parametrized using $S_{\alpha\sigma;11}$ and $S_{\alpha\sigma;12}$,
and we can explicitly reparametrize the kinetic variational parameters
as $n_{k\alpha\sigma;0}$ and $n_{k\alpha\sigma}$, where $n_{k\alpha\sigma;0}=0,1$
is the single particle density matrix of $\hat{K}_{2}$ and $n_{k\alpha\sigma}$
is the single particle density matrix of the SPD. It will be proven
that $n_{k\alpha\sigma;0}$ determines the Fermi surface of both the
interacting and non-interacting SPD, and therefore it will be useful
to define two regions of momentum space, denoted as $<$ or $>$,
where $<$ denotes the set of $k$ points with $n_{k\alpha\sigma;0}=1$
and $>$ indicates $n_{k\alpha\sigma;0}=0$; and we assume $\int dk=1$.
For each region $X\in\{<,>\}$ of a given spin orbital $\alpha\sigma$,
it will be useful to define the charge transfer $\Delta_{X\alpha\sigma}$
and charge fluctuation $\mathcal{A}_{X\alpha\sigma}$ as 

\begin{align}
 & \Delta_{X\alpha\sigma}=\int_{X}dk\left(n_{k\alpha\sigma;0}-n_{k\alpha\sigma}\right),\\
 & \mathcal{A}_{X\alpha\sigma}=\int_{X}dk\sqrt{n_{k\alpha\sigma}\left(1-n_{k\alpha\sigma}\right)},
\end{align}
which measure the influence of the local interaction on the given
spin orbital. Given that $\boldsymbol{g}'_{loc,\alpha\sigma;A}$ is
determined by $S_{\alpha\sigma;11}$, $S_{\alpha\sigma;12}$, $\left\{ n_{k\alpha\sigma}\right\} $,
and $\left\{ n_{k\alpha\sigma;0}\right\} $, the self-consistency
condition $\boldsymbol{g}_{loc;A}=\boldsymbol{g}'_{loc;A}$ becomes
three linear constraints on $n_{k\alpha\sigma;0}$ and $n_{k\alpha\sigma}$,
given as 
\begin{align}
\int n_{k\alpha\sigma;0}dk & =n_{\alpha\sigma}\left(\mathcal{G}_{12},u\right),\label{eq:linearc1}\\
\int n_{k\alpha\sigma}dk & =n_{\alpha\sigma}\left(\mathcal{G}_{12},u\right),\label{eq:linearc2}\\
\Delta_{<\alpha\sigma} & =\Delta_{\alpha\sigma}\left(\mathcal{G}_{12},u\right),\label{eq:linearc3}
\end{align}
where 
\begin{align}
 & n_{\alpha\sigma}\left(\mathcal{G}_{12},u\right)\equiv\left[\boldsymbol{g}_{loc;\alpha\sigma}\right]_{22},\\
 & \Delta_{\alpha\sigma}\left(\mathcal{G}_{12},u\right)\equiv\frac{S_{\alpha\sigma;12}}{S_{\alpha\sigma;11}}\left[\boldsymbol{g}_{loc;\alpha\sigma}\right]_{12}.
\end{align}
It should be noticed that the three constraints imply $\Delta_{<\alpha\sigma}=-\Delta_{>\alpha\sigma}$.
The three constraints have a clear interpretation. Equation \ref{eq:linearc1}
indicates that the local density of the non-interacting reference
system $\hat{K}_{2}$ is constrained to $\left[\boldsymbol{g}_{loc;\alpha\sigma}\right]_{22}$.
Using Equation \ref{eq:linearc2}, we see that $\left[\boldsymbol{g}_{loc;\alpha\sigma}\right]_{22}=\left[\boldsymbol{g}_{loc;\alpha\sigma}\right]_{33}$,
dictating that the Fermi volume is equal to the local density obtained
from the SPD. The result of these two constraints can be viewed as
the wave function analogue of the Luttinger theorem \cite{Luttinger19601153}.
The third constraint, Eq. \ref{eq:linearc3}, reveals how the local
interaction influences the density distribution. When the interacting
projector is close to the identity, $S_{\alpha\sigma;12}$ approaches
zero while the $S_{\alpha\sigma;11}$ and $\left[\boldsymbol{g}_{loc;\alpha\sigma}\right]_{12}$
approach a finite value, dictating that $\Delta_{\alpha\sigma}\left(\mathcal{G}_{12},u\right)$
approaches zero and therefore $n_{k\alpha\sigma}$ approaches $n_{k\alpha\sigma;0}$.
Alternatively, when the interacting projector deviates from the identity,
$\Delta_{\alpha\sigma}\left(\mathcal{G}_{12},u\right)$ increases
and imposes a deviation of $n_{k\alpha\sigma}$ away from $n_{k\alpha\sigma;0}$.
In summary, the first stage enforces self-consistency for the A block,
determining the kinetic energy. 

In the second stage, $\boldsymbol{g}'_{loc,\alpha\sigma}$ is fully
determined by $S_{\alpha\sigma;11}$, $S_{\alpha\sigma;12}$, $n_{\alpha\sigma}\left(\mathcal{G}_{12},u\right)$,
$\Delta_{\alpha\sigma}\left(\mathcal{G}_{12},u\right)$, $\mathcal{A_{<\alpha\sigma}}$,
and $\mathcal{A}_{>\alpha\sigma}$, therefore $\boldsymbol{g}_{loc}=\boldsymbol{g}'_{loc}$
can determine $\boldsymbol{\mathcal{G}}$ on the B, C, and D blocks,
which is derived in Section \ref{subsec:BCD-block}. In summary, the
self-consistency has been automatically satisfied and the local energy
determined.

In conclusion, we have an explicit functional form for the total energy
of the SPD parametrized by $\mathcal{G}_{12}$, $u$, and $n_{k\alpha\sigma}$,
given as
\begin{equation}
\sum_{\alpha\sigma}\int dk\epsilon_{k\alpha\sigma}n_{k\alpha\sigma}+E_{loc}\left(\mathcal{G}_{12},u,\mathcal{A}\right),\label{eq:energy_unminimized}
\end{equation}
where $\mathcal{A}=\{\mathcal{A_{<\alpha\sigma}},\mathcal{A_{>\alpha\sigma}}\}$
and $n_{k\alpha\sigma}$ is constrained by Equation \ref{eq:linearc2}
and \ref{eq:linearc3} and the volume of fermi sea is constrained
by Equation \ref{eq:linearc1}. The total energy has been expressed
as a functional of $\mathcal{G}_{12}$, $u$, $\{n_{k\alpha\sigma}$\},
and $\{n_{k\alpha\sigma;0}\}$. This algorithm can be viewed as a
nonlinear reparametrization of the original variational parameters
$|\Psi_{0}\rangle$, $u$, and $\{\gamma_{k\alpha\sigma}\}$, where
$\{n_{k\alpha\sigma;0}\}$ is a reparametrization of $|\Psi_{0}\rangle$,
$\{n_{k\alpha\sigma}\}$ is a reparametrization of part of $\{\gamma_{k\alpha\sigma}\}$,
and $\mathcal{G}_{12}$ can be viewed as a set of variational parameters
which reparametrizes the remaining part of $\{\gamma_{k\alpha\sigma}\}$
through condition \ref{eq:linearc3}. 

It should be noted that $\{n_{k\alpha\sigma}$\} only influences the
local interaction energy through $\mathcal{A}$, and is constrained
by $n_{\alpha\sigma}(\mathcal{G}_{12},u)$ and $\Delta_{\alpha\sigma}(\mathcal{G}_{12},u)$,
and therefore to find an optimized $n_{k\alpha\sigma}$ in the region
$X\in\{<,>\}$ of spin orbital $\alpha\sigma$, two Lagrange multipliers
$a_{X\alpha\sigma}$ and $b_{X\alpha\sigma}$ can be introduced 
\begin{align}
F_{X\alpha\sigma}=\int_{X}dk\Big( & \epsilon_{k\alpha\sigma}n_{k\alpha\sigma}-a_{X\alpha\sigma}n_{k\alpha\sigma}\nonumber \\
 & -b_{X\alpha\sigma}\sqrt{n_{k\alpha\sigma}\left(1-n_{k\alpha\sigma}\right)}\Big),
\end{align}
and we can solve for $n_{k\alpha\sigma}$ from $\frac{\delta F_{X\alpha\sigma}}{\delta n_{k\alpha\sigma}}\big|_{k\in X}=0\underline{}$,
resulting in
\begin{equation}
n_{k\alpha\sigma}\big|_{k\in X}=\frac{1}{2}\left(1+\frac{a_{X\alpha\sigma}-\epsilon_{k\alpha\sigma}}{\sqrt{\left(a_{X\alpha\sigma}-\epsilon_{k\alpha\sigma}\right)^{2}+b_{X\alpha\sigma}^{2}}}\right).\label{eq:dens_dist}
\end{equation}
Therefore, the true independent variational parameters for the algorithm
are $\mathcal{G}_{12}$, $u$, and $\bm{b}=\left\{ b_{X\alpha\sigma}\right\} $,
given that $\bm{a}=\{a_{X\alpha\sigma}\}$ can be determined as a
function of $\mathcal{G}_{12}$, $u$, and $\bm{b}$ through Eqs.
\ref{eq:linearc2} and \ref{eq:linearc3}. Finally, the ground state
energy can be determined as
\begin{align}
\mathcal{E}=\min_{\mathcal{G}_{12},u,\bm{b}} & \Big(\int dk\epsilon_{k\alpha\sigma}n_{k\alpha\sigma}\left(\bm{a},\bm{b}\right)+E_{loc}\left(\mathcal{G}_{12},u,\bm{b}\right)\Big),\label{eq:total_energy}
\end{align}
where the functional dependencies for $n_{k\alpha\sigma}\left(\bm{a},\bm{b}\right)$
are defined in Eq. \ref{eq:dens_dist} and $E_{loc}\left(\mathcal{G}_{12},u,\bm{b}\right)$
is detailed in the remaining sections. In this work, we used the Nelder-Mead
algorithm \cite{Nelder1965308} to perform the minimization in Eq.
\ref{eq:total_energy}, which is a gradient free algorithm. In some
cases, it may be preferable to solve a Hamiltonian with fixed density
$n_{\alpha\sigma}$, and this procedure is outlined in Appendix \ref{sec:app_density_constraint}. 

It is useful to give some practical guidelines for the efficiency
of the gauge constrained algorithm, which can roughly be broken down
into two factors. First, there is the cost of evaluating expectation
values under $\barhat[\rho]_{loc;i}$ (i.e. Eq. \ref{eq:local_expect}),
which will scale exponentially with the number of spin orbitals. Second,
there is the number of independent variational parameters, which scales
exponentially in the absence of symmetry. The first factor is roughly
independent of the symmetry of the Hamiltonian $\hat{H}$ which is
being solved, while the second factor strongly depends on the symmetry.
However, it is always possible to restrict the number of variational
parameters in order to control the computational cost of the second
factor, maintaining an upper bound for the total energy compared to
the full variational minimization. Therefore, there are numerous avenues
for engineering a minimal parametrization of the space of variational
parameters. In the present paper, we study the SU(2N$_{\textrm{orb}}$)
Hubbard model, where the high local symmetry results in a linear scaling
for the number of variational parameters, and therefore the first
factor completely dominates the computational cost. 

\subsection{Evaluating observables within the local A-block \label{subsec:A-block}}

Here we will elucidate why the block structure introduced in Eq. \ref{eq:block_matrix}
is the starting point for the gauge constrained algorithm. We begin
by explaining why $\bm{\mathcal{G}}_{\alpha\sigma;A}$ is the only
block that needs to be considered when determining $\boldsymbol{S}_{loc}$.
Given that $\barhat[P]$ only acts on the first and second integer
time step, $\boldsymbol{S}_{loc}$ only has nontrivial elements on
the A block, which are determined by $\bm{\mathcal{G}}_{\alpha\sigma;A}$
and $u$ (see Section V.B in Ref. \cite{Cheng2021195138} for further
background). Therefore, only the form of $\bm{\mathcal{G}}_{\alpha\sigma;A}$
needs to be specified to initiate the algorithm.

We previously demonstrated that the gauge freedom of the SPD allows
the following simple form \cite{Cheng2022205129}
\begin{equation}
\boldsymbol{\mathcal{G}}_{\alpha\sigma;A}=\begin{pmatrix}\frac{1}{2} & \mathcal{G}_{\alpha\sigma;12}\\
-\mathcal{G}_{\alpha\sigma;12} & \frac{1}{2}
\end{pmatrix},
\end{equation}
where $\mathcal{G}_{\alpha\sigma;\tau\tau'}=[\boldsymbol{\mathcal{G}}_{\alpha\sigma}]_{\tau\tau'}$
and $\mathcal{G}_{\alpha\sigma;12}\in[0,1/2]$. Since the $\boldsymbol{\mathcal{G}}_{\alpha\sigma;A}$
is completely determined, any observables within the local A block
can now be explicitly determined. For any operator $\barhat[O]$ local
to site $i$, the expectation value under $\barhat[\rho]_{loc;i}$
can be rewritten in terms of expectations values of the non-interacting
part of $\barhat[\rho]_{loc;i}$ as 
\begin{equation}
\langle\barhat[O]\rangle_{\barhat[\rho]_{loc;i}}=\frac{\langle\barhat[P]_{i}\barhat[O]\rangle_{\barhat[\rho]_{loc;i,0}}}{\langle\barhat[P]_{i}\rangle_{\barhat[\rho]_{loc;i,0}}},\label{eq:local_expect_values}
\end{equation}
where 
\begin{equation}
\barhat[\rho]_{loc;i,0}\equiv\exp(-\ln\left(\bm{\mathcal{G}}^{-1}-\boldsymbol{1}\right)^{T}\cdot\barhat[\bm{n}]_{i}),
\end{equation}
Using the form of $\barhat[P]_{i}$ in Eq. \ref{eq:local_int_proj},
we have
\begin{equation}
\langle\barhat[P]_{i}\barhat[O]\rangle_{\barhat[\rho]_{loc;i,0}}=u^{T}(\barhat[O])_{u}u,
\end{equation}
where $u=\left(u_{i1},\cdots,u_{iN_{\Gamma}}\right)^{T}$ is a $N_{\Gamma}$-element
real vector, $N_{\Gamma}$ is the number of local projectors, and
$(\barhat[O])_{u}$ is an $N_{\Gamma}\times N_{\Gamma}$ matrix with
elements 

\begin{equation}
[(\barhat[O])_{u}]_{\Gamma\Gamma'}=\langle\barhat[P]_{i\Gamma}^{\left(1\right)}\barhat[P]_{i\Gamma'}^{\left(2\right)}\barhat[O]\rangle_{\barhat[\rho]_{loc;i,0}}.\label{eq:Ogammagammap}
\end{equation}
It should be emphasized that the subscript in $(\barhat[O])_{u}$
solely indicates that this matrix and the vector $u$ are in the same
representation, and the elements of $(\barhat[O])_{u}$ defined in
Eq. \ref{eq:Ogammagammap} are not dependent on the values of $u_{i\Gamma}$;
a different representation which is useful for constraining the density
is presented in Appendix \ref{sec:app_density_constraint}. The expectation
value of $\barhat[O]$ under $\barhat[\rho]_{loc;i}$ is given as
\begin{equation}
\langle\barhat[O]\rangle_{\barhat[\rho]_{loc;i}}=\frac{u^{T}(\barhat[O])_{u}u}{u^{T}\left(\barhat[1]\right)_{u}u}.\label{eq:local_expect}
\end{equation}
For example, the local integer time Green's function can be computed
as
\begin{equation}
[\boldsymbol{g}_{loc;\alpha\sigma}]_{\tau\tau'}=\frac{u^{T}(\barhat[a]_{\alpha\sigma}^{\dagger(\tau)}\barhat[a]_{\alpha\sigma}^{(\tau')})_{u}u}{u^{T}\left(\barhat[1]\right)_{u}u}.\label{eq:gloc_expr}
\end{equation}
In the following, we present key formulas to evaluate equation \ref{eq:Ogammagammap}.
Given that we have restricted the SPD to be diagonal, the local projectors
can be chosen as \cite{Cheng2022205129}
\begin{equation}
\hat{P}_{i\Gamma}=\prod_{\alpha\sigma}\left(\delta_{\Gamma_{\alpha\sigma},0}(1-\hat{n}_{\alpha\sigma})+\delta_{\Gamma_{\alpha\sigma},1}\hat{n}_{\alpha\sigma}\right),\label{eq:SUN_projector}
\end{equation}
where $\Gamma_{\alpha\sigma}\in\{0,1\}$ and are determined from the
binary relation $\left(\Gamma_{1\uparrow}\Gamma_{1\downarrow}\dots\Gamma_{N_{\textrm{orb}}\uparrow}\Gamma_{N_{\textrm{orb}}\downarrow}\right)_{2}=\Gamma-1$.
The matrix elements of $\left(\barhat[1]\right)_{u}$ are given as
\begin{align}
 & \left[\left(\barhat[1]\right)_{u}\right]_{\Gamma\Gamma'}=\prod_{\alpha\sigma}p_{\alpha\sigma}\left(\Gamma_{\alpha\sigma},\Gamma'_{\alpha\sigma}\right),\label{eq:identity_in_compound}\\
 & p_{\alpha\sigma}\left(\Gamma_{\alpha\sigma},\Gamma_{\alpha\sigma}'\right)=\frac{1}{4}+\left(-1\right)^{\Gamma_{\alpha\sigma}+\Gamma'_{\alpha\sigma}}\mathcal{G}_{\alpha\sigma;12}^{2}.\label{eq:p_elements}
\end{align}
Single particle operators are evaluated as 
\begin{align}
[(\barhat[a]_{\alpha\sigma}^{\dagger\left(\tau\right)}\barhat[a]_{\alpha\sigma}^{\left(\tau'\right)})_{u}]_{\Gamma\Gamma'} & =g_{\alpha\sigma}^{\tau\tau'}\left(\Gamma_{\alpha\sigma},\Gamma'_{\alpha\sigma}\right)\nonumber \\
 & \times\prod_{\alpha'\sigma'\neq\alpha\sigma}p_{\alpha'\sigma'}\left(\Gamma_{\alpha'\sigma'},\Gamma'_{\alpha'\sigma'}\right),\label{eq:gloc_representation}
\end{align}
where
\begin{align}
 & g_{\alpha\sigma}^{\tau\tau'}\left(\Gamma_{\alpha\sigma},\Gamma'_{\alpha\sigma}\right)=p_{\alpha\sigma}\left(\Gamma_{\alpha\sigma},\Gamma'_{\alpha\sigma}\right)\mathcal{G}_{\alpha\sigma;\tau\tau'}+\left(-1\right)^{\Gamma_{\alpha\sigma}+\Gamma'_{\alpha\sigma}}\nonumber \\
 & \times\big((\frac{1}{2}\left(-1\right)^{\Gamma'_{\alpha\sigma}-1}\mathcal{G}_{\alpha\sigma;1\tau'}-\mathcal{G}_{\alpha\sigma;12}\mathcal{G}_{\alpha\sigma;2\tau'})(\delta_{1,\tau}-\mathcal{G}_{\alpha\sigma;\tau1})\nonumber \\
 & +(\frac{1}{2}\left(-1\right)^{\Gamma{}_{\alpha\sigma}-1}\mathcal{G}_{\alpha\sigma;2\tau'}+\mathcal{G}_{\alpha\sigma;12}\mathcal{G}_{\alpha\sigma;1\tau'})(\delta_{2,\tau}-\mathcal{G}_{\alpha\sigma;\tau2})\big).\label{eq:g_elements}
\end{align}
Any two particle correlation function of the below form are given
as 

\begin{align}
 & \left[\left(\barhat[a]_{\alpha_{1}\sigma_{1}}^{\dagger\left(\tau_{1}\right)}\barhat[a]_{\alpha_{1}\sigma_{1}}^{\left(\tau_{1}'\right)}\barhat[a]_{\alpha_{2}\sigma_{2}}^{\dagger\left(\tau_{2}\right)}\barhat[a]_{\alpha_{2}\sigma_{2}}^{\left(\tau_{2}'\right)}\right)_{u}\right]_{\Gamma\Gamma'}=g_{\alpha_{1}\sigma_{1}}^{\tau_{1}\tau_{1}'}\left(\Gamma_{\alpha_{1}\sigma_{1}},\Gamma'_{\alpha_{1}\sigma_{1}}\right)\nonumber \\
 & \times g_{\alpha_{2}\sigma_{2}}^{\tau_{2}\tau_{2}'}\left(\Gamma_{\alpha_{2}\sigma_{2}},\Gamma'_{\alpha_{2}\sigma_{2}}\right)\prod_{\alpha'\sigma'\neq\alpha_{1}\sigma_{1},\alpha_{2}\sigma_{2}}p_{\alpha'\sigma'}\left(\Gamma_{\alpha'\sigma'},\Gamma'_{\alpha'\sigma'}\right),
\end{align}
where $\alpha_{1}\sigma_{1}\neq\alpha_{2}\sigma_{2}$. In appendix
\ref{appendix:limitations}, we outline how to treat a general interacting
projector. In summary, we have provided explicit formulas for evaluating
local quantities up to the two particle level, which is sufficient
to execute the algorithm. It should be emphasized that these expressions
for local observables are valid outside of the A block, but require
complete knowledge of $\bm{\mathcal{G}}$ (e.g. see Eq. \ref{eq:g_elements}).

Normally, evaluating expectation values under $\barhat[\rho]_{loc;i}$
(i.e. Eq. \ref{eq:local_expect}) will be the rate limiting factor
in the SCDA, and given that $N_{\Gamma}$ scales exponentially with
the number of spin orbitals, the overall computational cost will scale
exponentially. There are two possible routes to mitigate this exponential
scaling. First, one could reduce the number of projectors, though
this must be done carefully as it will limit the variational freedom.
Second, one may use Monte Carlo to evaluate Eq. \ref{eq:local_expect}.

We now proceed to evaluate $\boldsymbol{S}_{loc,\alpha\sigma}$. Given
the choice of $\bm{\mathcal{G}}_{\alpha\sigma;A}$ and using equations
\ref{eq:identity_in_compound} and \ref{eq:gloc_representation},
we find that $\boldsymbol{g}_{\alpha\sigma;A}$ has the following
form
\begin{equation}
\boldsymbol{g}_{loc,\alpha\sigma;A}=\begin{pmatrix}n_{\alpha\sigma} & g_{\alpha\sigma;12}\\
-g_{\alpha\sigma;12} & n_{\alpha\sigma}
\end{pmatrix},
\end{equation}
where $n_{\alpha\sigma}$ and $g_{\alpha\sigma;12}$ are functions
of $\mathcal{G}_{12}$ and $u$. Given that the local interacting
projector only acts on the $A$ block, the discrete Dyson equation
simplifies to
\begin{equation}
\left(\boldsymbol{g}_{loc,\alpha\sigma;A}^{-1}-\boldsymbol{1}\right)=\left(\bm{\mathcal{G}}_{\alpha\sigma;A}^{-1}-\boldsymbol{1}\right)\boldsymbol{S}_{loc,\alpha\sigma;A},\label{eq:discrete_dyson_A_block}
\end{equation}
which yields an integer time self-energy of the form

\begin{equation}
\boldsymbol{S}_{loc,\alpha\sigma}=\left(\begin{array}{ccc}
S_{\alpha\sigma;11} & S_{\alpha\sigma;12} & 0\\
-S_{\alpha\sigma;12} & S_{\alpha\sigma;11} & 0\\
0 & 0 & 1
\end{array}\right),\label{eq:self}
\end{equation}
where
\begin{align}
S_{\alpha\sigma;11}= & \frac{1}{\left(4\mathcal{G}_{\alpha\sigma,1,2}^{2}+1\right)\left(g_{\alpha\sigma,1,2}^{2}+n_{\alpha\sigma}^{2}\right)}\nonumber \\
 & \times\Big(-g_{\alpha\sigma,1,2}^{2}+4g_{\alpha\sigma,1,2}\mathcal{G}_{\alpha\sigma,1,2}-n_{\alpha\sigma}^{2}+n_{\alpha\sigma}\nonumber \\
 & +4\mathcal{G}_{\alpha\sigma,1,2}^{2}\left(g_{\alpha\sigma,1,2}^{2}+\left(n_{\alpha\sigma}-1\right)n_{\alpha\sigma}\right)\Big),\label{eq:S11}\\
S_{\alpha\sigma;12}= & \frac{1}{\left(4\mathcal{G}_{\alpha\sigma,1,2}^{2}+1\right)\left(g_{\alpha\sigma,1,2}^{2}+n_{\alpha\sigma}^{2}\right)}\nonumber \\
 & \times\Big(g_{\alpha\sigma,1,2}\left(4\mathcal{G}_{\alpha\sigma,1,2}\left(\mathcal{G}_{\alpha\sigma,1,2}-g_{\alpha\sigma,1,2}\right)-1\right)\nonumber \\
 & -4\left(n_{\alpha\sigma}-1\right)n_{\alpha\sigma}\mathcal{G}_{\alpha\sigma,1,2}\Big).\label{eq:S12}
\end{align}
In summary, Eqns. \ref{eq:S11} and \ref{eq:S12} express the local
integer time self-energy as a function of $\mathcal{G}_{12}$ and
$u$.

\subsection{Parametrization of the integer time lattice Green's function and
self-consistency of the A-block\label{subsec:A-block_momentum_part}}

In the preceding section, we determined $\boldsymbol{S}_{loc}$, which
completely determines $\barhat[\rho]_{K}$ via Eq. \ref{eq:rhoK},
allowing the computation of $\bm{g}{}_{k\alpha\sigma}=\left\langle \barhat[\bm{n}]_{k\alpha\sigma}\right\rangle _{\barhat[\rho]_{K}}$.
We will demonstrate that $\bm{g}{}_{k\alpha\sigma}$ can be written
analytically in terms of $\gamma_{k\alpha\sigma}$, the expectation
value $n_{k\alpha\sigma;0}=\left\langle \hat{n}_{k\alpha\sigma}\right\rangle _{\hat{K}_{2}}$,
and $\boldsymbol{S}_{loc,\alpha\sigma}$. It is natural to reparametrize
$\gamma_{k\alpha\sigma}$ using $\lambda_{k\alpha\sigma}=\left\langle \hat{n}_{k\alpha\sigma}\right\rangle _{\hat{K}_{1}}=(1+\exp(-\gamma_{k\alpha\sigma}))^{-1}\in(0,1)$
\cite{Cheng2021195138}. In general, $\hat{K}_{2}$ can be a mixed
state, where $n_{k\alpha\sigma;0}\in\left[0,1\right]$, and an analytic
expression for $\bm{g}{}_{k\alpha\sigma}$ in terms of $\gamma_{k\alpha\sigma}$
and $n_{k\alpha\sigma;0}$ is given in the Appendix. At zero temperature
in the metallic phase, $\hat{K}_{2}$ will be a pure state after minimization
and $n_{k\alpha\sigma;0}$ is either zero or one. For the insulating
phase at zero temperature, $\bm{g}{}_{k\alpha\sigma}$ does not depend
on $n_{k\alpha\sigma;0}$, and therefore we are free to choose $n_{k\alpha\sigma;0}\in\left[0,1\right]$,
though for convenience we still choose zero or one. A general expression
for $\bm{g}{}_{k\alpha\sigma}$ is presented in Eq. S8 in Supplementary
Material \cite{supplementary}, which in the case of $n_{k\alpha\sigma;0}=1$
reduces to 

\begin{equation}
\boldsymbol{g}_{k\alpha\sigma}\big|_{n_{k\alpha\sigma;0}=1}=C^{-1}\begin{pmatrix}C & 0 & 0\\
-m_{21} & m_{22} & m_{23}\\
-m_{31} & -m_{32} & m_{22}
\end{pmatrix},
\end{equation}
where

\begin{align}
 & m_{21}=\left(\lambda_{k\alpha\sigma}-1\right){}^{2}S_{\alpha\sigma;11},\\
 & m_{22}=\lambda_{k\alpha\sigma}^{2},\\
 & m_{23}=\left(1-\lambda_{k\alpha\sigma}\right)\lambda_{k\alpha\sigma},\\
 & m_{31}=\left(1-\lambda_{k\alpha\sigma}\right)\lambda_{k\alpha\sigma}S_{\alpha\sigma;11},\\
 & m_{32}=\left(1-\lambda_{k\alpha\sigma}\right)\lambda_{k\alpha\sigma}S_{\alpha\sigma;12},\\
 & C=\lambda_{k\alpha\sigma}^{2}+\left(\lambda_{k\alpha\sigma}-1\right){}^{2}S_{\alpha\sigma;12}.
\end{align}
Furthermore, it is natural to reparametrize $\lambda_{k\alpha\sigma}$
using $n_{k\alpha\sigma}=\left[\boldsymbol{g}_{k\alpha\sigma}\right]_{33}=\langle\barhat[n]_{k\alpha\sigma}^{(3)}\rangle_{\barhat[\rho]_{K}}$,
which is the physical density distribution $\left\langle \hat{n}_{k\alpha\sigma}\right\rangle _{\spd}$
within the SCDA, as 
\begin{equation}
\lambda_{k\alpha\sigma}=\frac{n_{k\alpha\sigma}S_{\alpha\sigma;12}}{n_{k\alpha\sigma}S_{\alpha\sigma;12}+\sqrt{\left(1-n_{k\alpha\sigma}\right)n_{k\alpha\sigma}S_{\alpha\sigma;12}}},
\end{equation}
resulting in 
\begin{align}
 & \boldsymbol{g}_{k\alpha\sigma}\Big|_{n_{k\alpha\sigma;0}=1}\nonumber \\
 & =\left(\begin{array}{ccc}
1 & 0 & 0\\
-\frac{S_{\alpha\sigma;11}}{S_{\alpha\sigma;12}}\left(1-n_{k\alpha\sigma}\right) & n_{k\alpha\sigma} & \frac{A}{\sqrt{S_{\alpha\sigma;12}}}\\
-\frac{S_{\alpha\sigma;11}}{\sqrt{S_{\alpha\sigma;12}}}A & -\sqrt{S_{\alpha\sigma;12}}A & n_{k\alpha\sigma}
\end{array}\right),
\end{align}
where $A=\sqrt{\left(1-n_{k\alpha\sigma}\right)n_{k\alpha\sigma}}$.
For the case of $n_{k\alpha\sigma;0}=0$, we have

\begin{equation}
\boldsymbol{g}_{k\alpha\sigma}\Big|_{n_{k\alpha\sigma;0}=0}=C^{-1}\left(\begin{array}{ccc}
0 & m_{12} & m_{13}\\
0 & m_{22} & m_{23}\\
0 & -m_{32} & m_{22}
\end{array}\right),
\end{equation}
where

\begin{align}
 & m_{12}=\lambda_{k\alpha\sigma}^{2}S_{\alpha\sigma;11},\\
 & m_{13}=\left(1-\lambda_{k\alpha\sigma}\right)\lambda_{k\alpha\sigma}S_{\alpha\sigma;11},\\
 & m_{22}=\lambda_{k\alpha\sigma}^{2}S_{\alpha\sigma;12},\\
 & m_{23}=\left(1-\lambda_{k\alpha\sigma}\right)\lambda_{k\alpha\sigma}S_{\alpha\sigma;12},\\
 & m_{32}=\left(1-\lambda_{k\alpha\sigma}\right)\lambda_{k\alpha\sigma}\left(S_{\alpha\sigma;11}^{2}+S_{\alpha\sigma;12}^{2}\right),\\
 & C=\left(\lambda_{k\alpha\sigma}-1\right){}^{2}\left(S_{\alpha\sigma;11}^{2}+S_{\alpha\sigma;12}^{2}\right)+\lambda_{k\alpha\sigma}^{2}S_{\alpha\sigma;12}.
\end{align}
We can similarly reparametrize $\lambda_{k\alpha\sigma}$ in terms
of $n_{k\alpha\sigma}$ as

\begin{equation}
\lambda_{k\alpha\sigma}=\frac{\sqrt{n_{k\alpha\sigma}}S_{\alpha\sigma}}{\sqrt{n_{k\alpha\sigma}}S_{\alpha\sigma}+\sqrt{(1-n_{k\alpha\sigma})S_{\alpha\sigma;12}}},
\end{equation}
where $S_{\alpha\sigma}=\sqrt{S_{\alpha\sigma;11}^{2}+S_{\alpha\sigma;12}^{2}}$,
yielding

\begin{equation}
\boldsymbol{g}_{k\alpha\sigma}\Big|_{n_{k\alpha\sigma;0}=0}=\left(\begin{array}{ccc}
0 & \frac{S_{\alpha\sigma;11}}{S_{\alpha\sigma;12}}n_{k\alpha\sigma} & \frac{S_{\alpha\sigma;11}}{\sqrt{S_{\alpha\sigma;12}}S_{\alpha\sigma}}A\\
0 & n_{k\alpha\sigma} & \frac{\sqrt{S_{\alpha\sigma;12}}}{S_{\alpha\sigma}}A\\
0 & -\frac{S_{\alpha\sigma}}{\sqrt{S_{\alpha\sigma;12}}}A & n_{k\alpha\sigma}
\end{array}\right),
\end{equation}
where $A=\sqrt{\left(1-n_{k\alpha\sigma}\right)n_{k\alpha\sigma}}$.
The local integer time Green's function can now be constructed as
an average over the Brillouin zone as
\begin{equation}
\boldsymbol{g}'_{loc,\alpha\sigma}=\langle\barhat[\boldsymbol{n}]_{i\alpha\sigma}\rangle_{\barhat[\rho]_{K}}=\int dk\boldsymbol{g}_{k\alpha\sigma},
\end{equation}
using the convention $\int dk=1$. 

Using the self-consistency condition on the $A$ block, 
\begin{equation}
\boldsymbol{g}'_{loc,\alpha\sigma;A}=\boldsymbol{g}{}_{loc,\alpha\sigma;A},
\end{equation}
we can determine the resulting constraints on $n_{k\alpha\sigma;0}$,
$n_{k\alpha\sigma}$, $u$, and $\mathcal{G}_{\alpha\sigma;12}$.
There are four constraining equations from the four corresponding
entries of the $A$ block, but only three of them are independent.
The first constraint is $[\boldsymbol{g}'_{loc,\alpha\sigma}]_{11}=[\boldsymbol{g}_{loc,\alpha\sigma}]_{11}$,
which yields 
\begin{equation}
\int dkn_{k\alpha\sigma;0}=\int_{<}dk=n_{\alpha\sigma}\left(\mathcal{G}_{12},u\right),\label{eq:condition_first}
\end{equation}
where $n_{\alpha\sigma}\left(\mathcal{G}_{12},u\right)=[\boldsymbol{g}_{loc,\alpha\sigma}]_{11}=[\boldsymbol{g}_{loc,\alpha\sigma}]_{22}$,
the symbol $<$ denotes the region where $n_{k\alpha\sigma;0}=1$,
while $>$ denotes the region where $n_{k\alpha\sigma;0}=0$. The
first constraint requires that $|\Psi_{0}\rangle$ has the same density
as given by $n_{\alpha\sigma}\left(\mathcal{G}_{12},u\right)$, and
we refer to this as the fermi volume constraint. The second constraint
is $[\boldsymbol{g}'_{loc,\alpha\sigma}]_{22}=[\boldsymbol{g}_{loc,\alpha\sigma}]_{22}$,
which yields the density constraint
\begin{equation}
\int dkn_{k\alpha\sigma}=\int_{<}dkn_{k\alpha\sigma}+\int_{>}dkn_{k\alpha\sigma}=n_{\alpha\sigma}\left(\mathcal{G}_{12},u\right).\label{eq:condition_second}
\end{equation}
The third constraint is $[\boldsymbol{g}'_{loc,\alpha\sigma}]_{12}=[\boldsymbol{g}_{loc,\alpha\sigma}]_{12}$
, which yields 
\begin{equation}
\int_{>}dkn_{k\alpha\sigma}=\Delta_{\alpha\sigma}\left(\mathcal{G}_{12},u\right),\label{eq:condition_third}
\end{equation}
where $\Delta_{\alpha\sigma}\left(\mathcal{G}_{12},u\right)\equiv\frac{S_{\alpha\sigma;12}}{S_{\alpha\sigma;11}}\left[\boldsymbol{g}_{loc;\alpha\sigma}\right]_{12}.$
The fourth constraint is $[\boldsymbol{g}'_{loc,\alpha\sigma}]_{21}=[\boldsymbol{g}_{loc,\alpha\sigma}]_{21}$
, which yields 
\begin{equation}
\int_{<}dk\left(1-n_{k\alpha\sigma}\right)=\Delta_{\alpha\sigma}\left(\mathcal{G}_{12},u\right).\label{eq:condition_fourth}
\end{equation}
The third and fourth constraint are identical as long as the first
and second constraint are satisfied, 
\begin{equation}
\int_{<}dk\left(1-n_{k\alpha\sigma}\right)=\int_{>}dkn_{k\alpha\sigma}=\Delta_{\alpha\sigma}\left(\mathcal{G}_{12},u\right),\label{eq:condition_combined}
\end{equation}
which we refer to as the charge transfer constraint. 

We now discuss how to satisfy these three constraints, using constraints
on $n_{k\alpha\sigma;0}$ and $n_{k\alpha\sigma}$. One can start
with arbitrary $u$ and $\mathcal{G}_{\alpha\sigma;12}\in[0,1/2]$,
which yields some $n_{\alpha\sigma}\left(\mathcal{G}_{12},u\right)$
that determines the fermi volume and $n_{k\alpha\sigma;0}$. Furthermore,
one must choose $n_{k\alpha\sigma}$ such that Eqns. \ref{eq:condition_second}
and \ref{eq:condition_combined} are satisfied. To simplify the expression
for $\boldsymbol{g}'_{loc}$, it is useful to define the following
quantities 
\begin{align}
 & \Delta_{<\alpha\sigma}=\int_{<}dk\left(n_{k\alpha\sigma;0}-n_{k\alpha\sigma}\right)=\int_{<}dk\left(1-n_{k\alpha\sigma}\right),\\
 & \Delta_{>\alpha\sigma}=\int_{>}dk(n_{k\alpha\sigma;0}-n_{k\alpha\sigma})=-\int_{>}dkn_{k\alpha\sigma},\\
 & \mathcal{A}_{<\alpha\sigma}=\int_{<}dk\sqrt{n_{k\alpha\sigma}\left(1-n_{k\alpha\sigma}\right)},\\
 & \mathcal{A}_{>\alpha\sigma}=\int_{>}dk\sqrt{n_{k\alpha\sigma}\left(1-n_{k\alpha\sigma}\right)}.
\end{align}
Using equations \ref{eq:condition_first} and \ref{eq:condition_second},
we have $\Delta_{<\alpha\sigma}=-\Delta_{>\alpha\sigma}$. Equations
\ref{eq:condition_first} and \ref{eq:condition_second} are treated
as independent conditions, and equations \ref{eq:condition_third}
and \ref{eq:condition_fourth} become the single condition given in
Eq. \ref{eq:condition_combined}. For convenience, we define 
\begin{align}
\Delta_{\alpha\sigma} & \equiv\Delta_{<\alpha\sigma}=-\Delta_{>\alpha\sigma},\\
n_{\alpha\sigma} & \equiv\int_{<}dk=\int dkn_{k\alpha\sigma},
\end{align}
which should not be confused with the corresponding quantities $\Delta_{\alpha\sigma}\left(\mathcal{G}_{12},u\right)$
and $n_{\alpha\sigma}\left(\mathcal{G}_{12},u\right)$ determined
from the local discrete action. The quantity $\Delta_{\alpha\sigma}$
measures the total charge transfer generated by the projector $\hat{K}_{1}\hat{P}_{1}$
across the fermi surface determined by $|\Psi_{0}\rangle$, and is
uniquely determined from $\barhat[\rho]_{loc;i}$. The non-interacting
case yields $\Delta_{\alpha\sigma}=0$, while the strong coupling
limit of the Mott insulating phase yields $n_{k\alpha\sigma}=n_{\alpha\sigma}$
and $\Delta_{\alpha\sigma}=n_{\alpha\sigma}\left(1-n_{\alpha\sigma}\right)$.
Once $n_{k\alpha\sigma}$ and $n_{k\alpha\sigma;0}$ have been constrained
by equations \ref{eq:condition_first}, \ref{eq:condition_second},
and \ref{eq:condition_combined}, the total kinetic energy can be
evaluated as 
\begin{equation}
K=\sum_{\alpha\sigma}\int dk\epsilon_{k\alpha\sigma}n_{k\alpha\sigma}.
\end{equation}

We now discuss the properties of $\mathcal{A}_{<\alpha\sigma}$ and
$\mathcal{A}_{>\alpha\sigma}$, which are relevant for evaluating
the $B$, $C$, and $D$ blocks of $\boldsymbol{g}_{loc;\alpha\sigma}'$.
The non-interacting case yields $\mathcal{A}_{<\alpha\sigma}=\mathcal{A}_{>\alpha\sigma}=0$,
while for a given $n_{\alpha\sigma}$ and $\Delta_{\alpha\sigma}$
the maximum value of $\mathcal{A}_{<\alpha\sigma}$ is reached when
$n_{k\alpha\sigma}|{}_{k\in<}=1-\Delta_{\alpha\sigma}/n_{\alpha\sigma}$,
and therefore $\mathcal{A}_{<\alpha\sigma}\in[0,\sqrt{(n_{\alpha\sigma}-\Delta_{\alpha\sigma})\Delta_{\alpha\sigma}}]$.
Similarly, the maximum of $\mathcal{A}_{>\alpha\sigma}$ is reached
when $n_{k\alpha\sigma}|{}_{k\in>}=\Delta_{\alpha\sigma}/\left(1-n_{\alpha\sigma}\right)$
and therefore $\mathcal{A}_{>\alpha\sigma}\in[0,\sqrt{(1-n_{\alpha\sigma}-\Delta_{\alpha\sigma})\Delta_{\alpha\sigma}}]$.
Using these definitions, we have 
\begin{align}
 & \int_{<}dk\boldsymbol{g}_{k\alpha\sigma}\nonumber \\
 & =\left(\begin{array}{ccc}
n_{\alpha\sigma} & 0 & 0\\
-\frac{S_{\alpha\sigma;11}}{S_{\alpha\sigma;12}}\Delta_{\alpha\sigma} & n_{\alpha\sigma}-\Delta_{\alpha\sigma} & \frac{\mathcal{A}_{<,\alpha\sigma}}{\sqrt{S_{\alpha\sigma;12}}}\\
-\frac{S_{\alpha\sigma;11}}{\sqrt{S_{\alpha\sigma;12}}}\mathcal{A}_{<,\alpha\sigma} & -\sqrt{S_{\alpha\sigma;12}}\mathcal{A}_{<,\alpha\sigma} & n_{\alpha\sigma}-\Delta_{\alpha\sigma}
\end{array}\right),
\end{align}
and

\begin{align}
 & \int_{>}dk\boldsymbol{g}_{k\alpha\sigma}\nonumber \\
 & =\left(\begin{array}{ccc}
0 & \frac{S_{\alpha\sigma;11}}{S_{\alpha\sigma;12}}\Delta_{\alpha\sigma} & \frac{S_{\alpha\sigma;11}}{\sqrt{S_{\alpha\sigma;12}}S_{\alpha\sigma}}\mathcal{A}_{>,\alpha\sigma}\\
0 & \Delta_{\alpha\sigma} & \frac{\sqrt{S_{\alpha\sigma;12}}}{S_{\alpha\sigma}}\mathcal{A}_{>,\alpha\sigma}\\
0 & -\frac{S_{\alpha\sigma}}{\sqrt{S_{\alpha\sigma;12}}}\mathcal{A}_{>,\alpha\sigma} & \Delta_{\alpha\sigma}
\end{array}\right),
\end{align}
yielding the local lattice Green's function 
\begin{equation}
\boldsymbol{g}'_{loc,\alpha\sigma}=\left(\begin{array}{ccc}
n_{\alpha\sigma} & \frac{S_{\alpha\sigma;11}}{S_{\alpha\sigma;12}}\Delta_{\alpha\sigma} & g_{13}\\
-\frac{S_{\alpha\sigma;11}}{S_{\alpha\sigma;12}}\Delta_{\alpha\sigma} & n_{\alpha\sigma} & g_{23}\\
g_{31} & g_{32} & n_{\alpha\sigma}
\end{array}\right),
\end{equation}
where

\begin{align}
 & g_{13}=\frac{S_{\alpha\sigma;11}}{\sqrt{S_{\alpha\sigma;12}}S_{\alpha\sigma}}\mathcal{A}_{>,\alpha\sigma},\\
 & g_{23}=\frac{1}{\sqrt{S_{\alpha\sigma;12}}}\mathcal{A}_{<,\alpha\sigma}+\frac{\sqrt{S_{\alpha\sigma;12}}}{S_{\alpha\sigma}}\mathcal{A}_{>,\alpha\sigma},\\
 & g_{31}=-\frac{S_{\alpha\sigma;11}}{\sqrt{S_{\alpha\sigma;12}}}\mathcal{A}_{<,\alpha\sigma},\\
 & g_{32}=-\sqrt{S_{\alpha\sigma;12}}\mathcal{A}_{<,\alpha\sigma}-\frac{S_{\alpha\sigma}}{\sqrt{S_{\alpha\sigma;12}}}\mathcal{A}_{>,\alpha\sigma}.
\end{align}
The above equations provide explicit expressions for all blocks of
$\boldsymbol{g}_{loc;\alpha\sigma}'$.

\subsection{Determining the $B$, $C$, and $D$ blocks of $\boldsymbol{\mathcal{G}}$
and evaluating the total energy\label{subsec:BCD-block}}

Assuming that $\boldsymbol{S}_{loc}$ and $\boldsymbol{g}_{loc}'$
have been completely determined, the self-consistency condition $\boldsymbol{g}_{loc}=\boldsymbol{g}_{loc}'$
can be used to determine the remaining blocks of $\boldsymbol{\mathcal{G}}$
via the discrete Dyson equation as

\begin{equation}
\boldsymbol{\mathcal{G}}=\left(\boldsymbol{1}+\left(\boldsymbol{g}_{loc}^{-1}-\boldsymbol{1}\right)\boldsymbol{S}_{loc}^{-1}\right)^{-1},
\end{equation}
which yields

\begin{align}
 & \boldsymbol{\mathcal{G}}_{B}=\boldsymbol{\mathcal{G}}_{A}\boldsymbol{g}_{loc;A}^{-1}\boldsymbol{g}_{loc;B},\\
 & \boldsymbol{\mathcal{G}}_{C}=\boldsymbol{g}_{loc;C}\left(\boldsymbol{1}-\boldsymbol{g}_{loc;A}\right)^{-1}\left(\boldsymbol{1}-\boldsymbol{\mathcal{G}}_{A}\right),\\
 & \boldsymbol{\mathcal{G}}_{D}=\boldsymbol{g}_{loc;D}\nonumber \\
 & +\boldsymbol{g}_{loc;C}\left(\boldsymbol{1}-\boldsymbol{g}_{loc;A}\right)^{-1}\left(\boldsymbol{g}_{loc;A}-\boldsymbol{\mathcal{G}}_{A}\right)\boldsymbol{g}_{loc;A}^{-1}\boldsymbol{g}_{loc;B}.
\end{align}
The individual matrix elements of $\boldsymbol{\mathcal{G}}$ are
provided in Supplementary Material \cite{supplementary}.

We now proceed to compute the total energy. Having completely determined
$\boldsymbol{\mathcal{G}}$, the local interaction energy can be computed
using Eq. \ref{eq:local_expect} as
\begin{equation}
E_{loc}=\frac{u^{T}\left(\barhat[H]_{loc}^{(3)}\right)_{u}u}{u^{T}\left(\barhat[1]\right)_{u}u},\label{eq:Eloc}
\end{equation}
where the matrix $(\barhat[H]_{loc}^{(3)})_{u}$ depends on $\boldsymbol{\mathcal{G}}$
and the matrix $\left(\barhat[1]\right)_{u}$ depends on $\boldsymbol{\mathcal{G}}_{A}$.
It should be emphasized that the density distribution $n_{k\alpha\sigma}$
will influence $E_{loc}$ through $\boldsymbol{\mathcal{G}}_{B}$and
$\boldsymbol{\mathcal{G}}_{C}$.

\section{Results for SU(2N$_{\textrm{orb}}$) Hubbard model\label{sec:Results-for-SU(2N)}}

\begin{figure}[tbph]
\includegraphics[width=0.96\columnwidth]{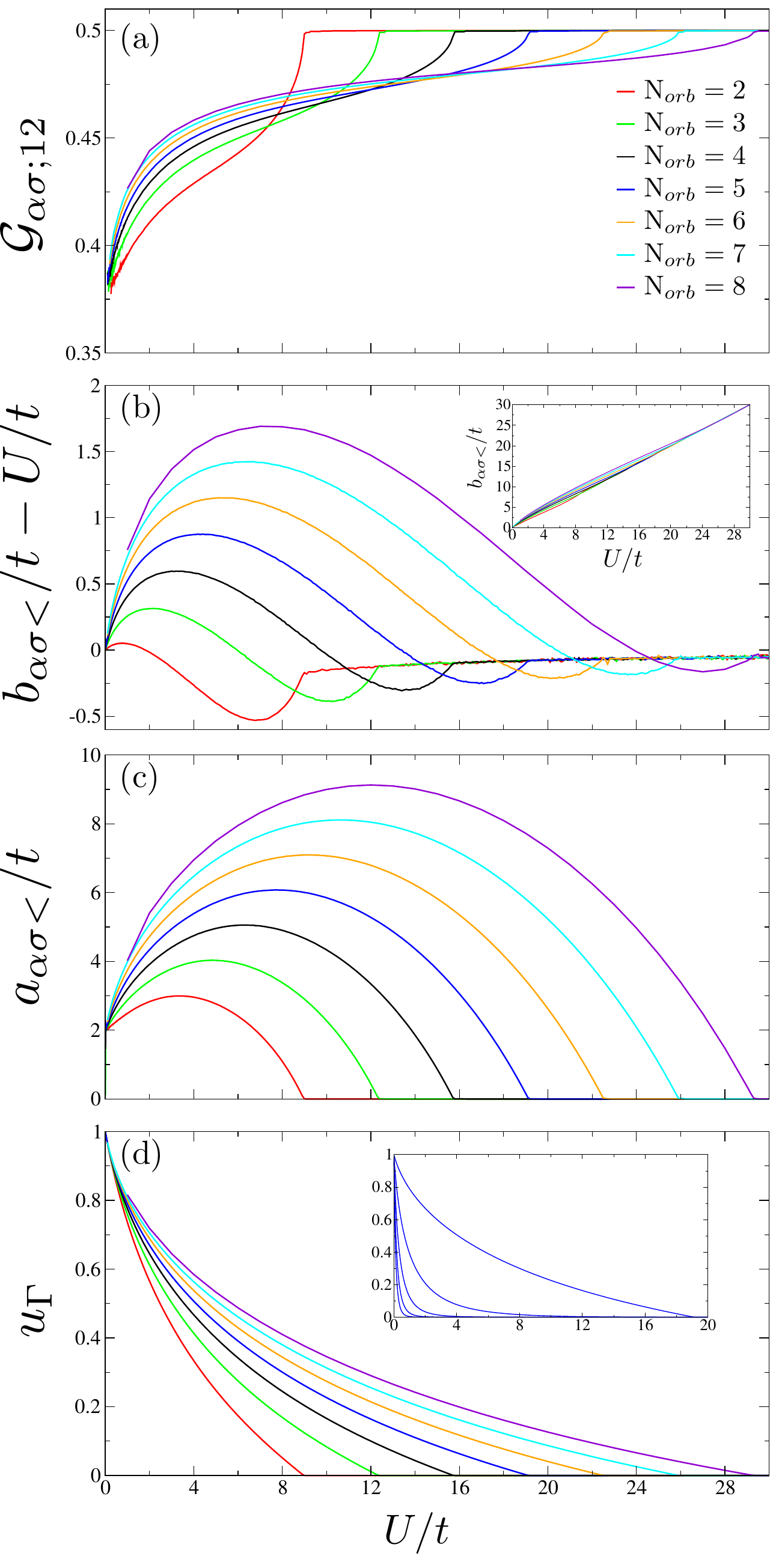}

\caption{\label{fig:Variational-Parameters}Optimized variational parameters
of VDAT within the SCDA at $\discn=3$ for the SU(2N$_{\textrm{orb}}$)
Hubbard model on the $d=\infty$ Bethe lattice at half filling and
$T=0$, with N$_{\textrm{orb}}=2-8$. The variational parameters $\mathcal{G}_{\alpha\sigma;12}$,
$b_{\alpha\sigma<}$, and $u_{\Gamma}$ for $\Gamma$ with particle
number N$_{\textrm{orb}}-1$ are shown in panels $a$, $b$, and $d$,
respectively, while the dependent parameters $a_{\alpha\sigma<}$
are shown in panel $c$. The inset of panel $d$ shows all $u_{\Gamma}$
for N$_{\textrm{orb}}=5$, and the value monotonically decreases as
the particle number of $\Gamma$ decreases from four to zero.}
\end{figure}

We now proceed to illustrate the gauge constrained implementation
of VDAT within the SCDA at $\discn=3$. We choose to study the SU(2N$_{\textrm{orb}}$)
Hubbard model, where $\hat{H}_{loc;i}=U\sum_{\ell<\ell'}\hat{n}_{i\ell}\hat{n}_{i\ell'}$
and $\ell=1,\dots,2\textmd{N}_{\textrm{orb}}$, in order to showcase
the advantages of VDAT over DMFT for obtaining zero temperature results
at large N$_{\textrm{orb}}$. It should be noted that the SU(2N$_{\textrm{orb}}$)
symmetry can be exploited when evaluating the expectation values of
local observables (i.e. Eq. \ref{eq:local_expect}), but for purposes
of benchmarking the computational cost, we utilized the general algorithm
which does not exploit the SU(2N$_{\textrm{orb}}$) symmetry. Therefore,
for a single evaluation of the SPD, the computational cost is the
same as the general case. To provide a rough idea of the computational
cost on a typical single processor core, when $\textrm{N}{}_{\textrm{orb}}\le3$
the cost of a single evaluation of the SPD is $10^{-4}$ seconds.
For $\textrm{N}{}_{\textrm{orb}}>3$, the computational cost scales
exponentially, requiring $10^{-3}$, 0.02, 0.1, and 3 seconds for
$\textrm{N}{}_{\textrm{orb}}$ of 4, 5, 6, and 7, respectively. The
timing difference between $\discn=2$ and $\discn=3$ is negligible
in all the aforementioned cases. We study the SU(2N$_{\textrm{orb}}$)
Hubbard model on the $d=\infty$ Bethe lattice for N$_{\textrm{orb}}$=2-8
at half filling and 2N$_{\textrm{orb}}$=5 for all fillings. At half
filling for N$_{\textrm{orb}}$=2,4 we compare to a zero temperature
extrapolation of published DMFT results using a quantum Monte-Carlo
(QMC) impurity solver \cite{Blumer2013085115}, while for 2N$_{\textrm{orb}}$=5
we compare to published DMFT results using the numerical renormalization
group (NRG) impurity solver \cite{Lee2018165143}. All VDAT results
in this paper are generated by a Julia implementation of the gauge
constrained algorithm, which can treat general density-density interactions,
and is available at \cite{Cheng_VDATN3multi_jl_2023}.

\begin{figure}[tbph]
\includegraphics[width=0.98\columnwidth]{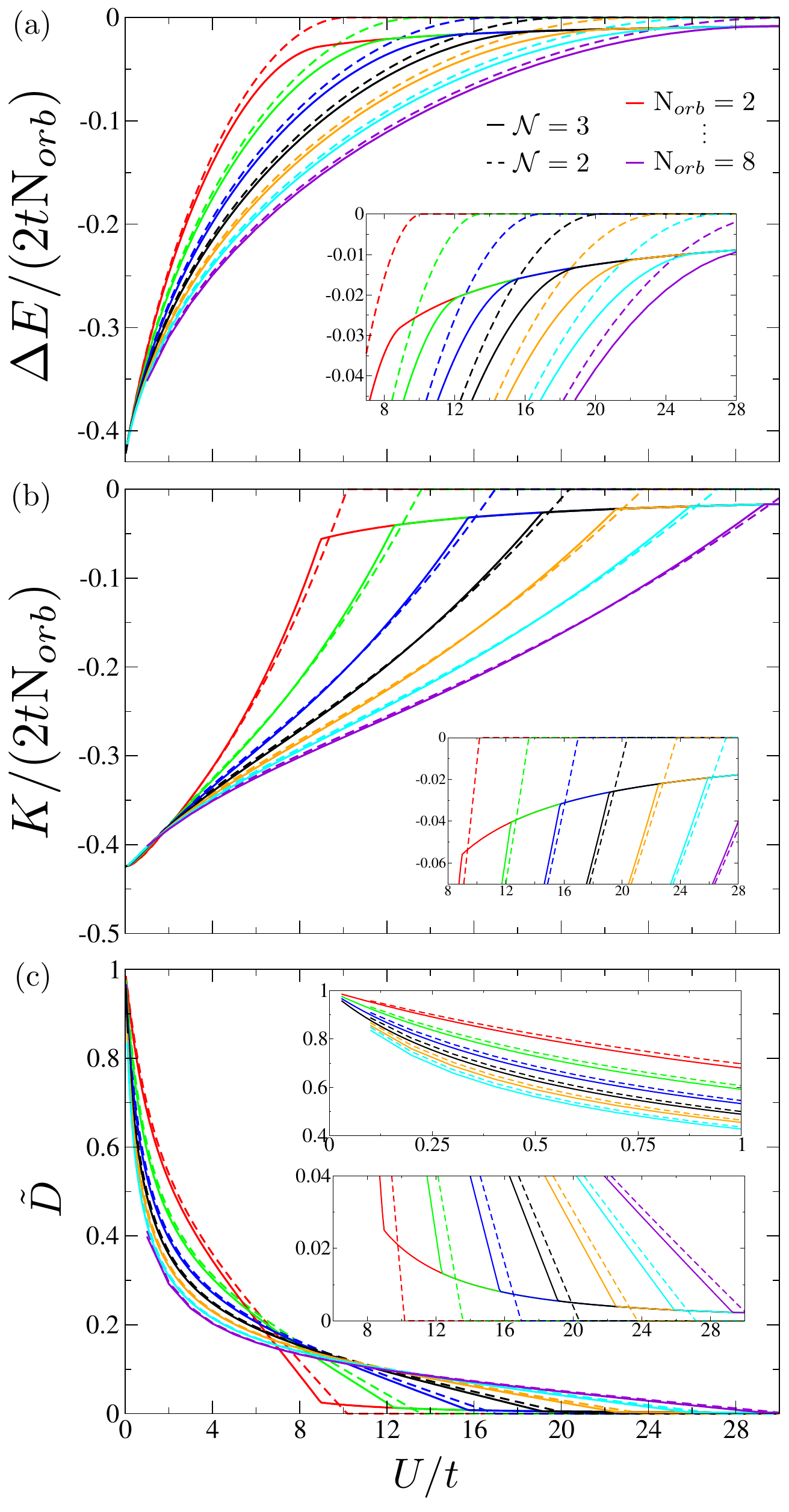}

\caption{Static observables computed from VDAT within the SCDA at $\discn=2$
(dashed lines) and $\discn=3$ (solid lines) for the SU(2N$_{\textrm{orb}}$)
Hubbard model on the $d=\infty$ Bethe lattice at half filling and
$T=0$, with N$_{\textrm{orb}}=2-8$. The legend colors are identical
to Figure \ref{fig:Variational-Parameters}. ($a$) Total energy difference
per spin orbital vs. $U/t$, where $\Delta E=E(t,U)-E(0,U)$. ($b$)
Kinetic energy per spin orbital vs. $U/t$. ($c$) Scaled double occupancy
vs. $U/t$, where $\tilde{D}=(D-D_{at})/(D_{0}-D_{at})$; $D$, $D_{0}$,
and $D_{at}$ are the double occupancies at the given $U/t$, the
non-interacting value, and the atomic value, respectively. \label{fig:energies}}
\end{figure}

We begin by considering the case of half filling for N$_{\textrm{orb}}$=2-8,
exceeding the number of orbitals that would be encountered for a $d$
or $f$ electron manifold relevant to strongly correlated electron
materials. The basic aspects of the SU(2N$_{\textrm{orb}}$) Hubbard
model in $d=\infty$ are well understood: there is a metal-insulator
transition (MIT) at a critical value of $U$, and this transition
value increases with N$_{\textrm{orb}}$. The signature of the MIT
can be seen in the variational parameters of the SPD, and therefore
it is instructive to examine $\mathcal{G}_{\alpha\sigma;12}$, $b_{\alpha\sigma<}$,
and the local variational parameters $u=\{u_{\Gamma}\}$ (see Figure
\ref{fig:Variational-Parameters}). It is also useful to explore the
intermediate parameters $a_{\alpha\sigma<}$, which can be determined
from the variational parameters. Because of the SU(2N$_{\textrm{orb}}$)
symmetry, the $\mathcal{G}_{\alpha\sigma;12}$, $b_{\alpha\sigma<}$,
and $a_{\alpha\sigma<}$ are independent of $\alpha\sigma$, while
$u_{\Gamma}$ only depends on the particle number of $\Gamma$. Furthermore,
particle-hole symmetry at half filling dictates that $b_{\alpha\sigma<}=b_{\alpha\sigma>}$,
$a_{\alpha\sigma<}=-a_{\alpha\sigma>}$, and $u_{\Gamma}=u_{\Gamma'}$
if the sum of the particle numbers of $\Gamma$ and $\Gamma'$ is
2N$_{\textrm{orb}}$. Considering $\mathcal{G}_{\alpha\sigma;12}$,
it begins at $U=0$ in the metallic phase with a value of roughly
0.37 and monotonically increases to 0.5 at the MIT, and is fixed at
0.5 in the insulating regime (see Figure \ref{fig:Variational-Parameters}a).
Increasing the number of orbitals from N$_{\textrm{orb}}$=2-8 has
several effects. In the small $U$ regime, $\mathcal{G}_{\alpha\sigma;12}$
increases with N$_{\textrm{orb}}$, enhancing the electronic correlations
for larger N$_{\textrm{orb}}$. Alternatively, increasing N$_{\textrm{orb}}$
will increase the critical value of $U$ for the metal-insulator transition.
The variational parameter $b_{\alpha\sigma<}$ turns out to be approximately
equal to $U$, and therefore we also plot $b_{\alpha\sigma<}-U$ (see
Figure \ref{fig:Variational-Parameters}b). The intermediate parameter
$a_{\alpha\sigma<}$, which can be computed from the variational parameters,
has a value of $2t$ for $U=0$, independent of N$_{\textrm{orb}}$,
and goes to zero in the insulating phase (see Figure \ref{fig:Variational-Parameters}c).
The latter can be appreciated from Eq. \ref{eq:dens_dist}, which
dictates that the quasiparticle weight becomes zero when $a_{\alpha\sigma<}=a_{\alpha\sigma>}$
and $b_{\alpha\sigma<}=b_{\alpha\sigma>}$. For a given N$_{\textrm{orb}}$,
there is an arbitrary coefficient for the interacting projector which
can be fixed by choosing $u_{\Gamma}=1$ when the particle number
of $\Gamma$ is N$_{\textrm{orb}}$. Therefore, there are N$_{\textrm{orb}}$
independent local variational parameters $u_{\Gamma}$ for $\Gamma$
with particle number equal to $0,\dots,\textrm{N}_{\textrm{orb}}-1$.
For a given N$_{\textrm{orb}}$, we plot the $u_{\Gamma}$ where the
particle number for $\Gamma$ is N$_{\textrm{orb}}$-1 (see Figure
\ref{fig:Variational-Parameters}d), in addition to all $u_{\Gamma}$
for $\textrm{N}_{\textrm{orb}}=5$. The $u_{\Gamma}$ for $\Gamma$
with particle number N$_{\textrm{orb}}$-1 goes to zero in the insulating
phase.

We now consider the total energy, where we explore both $\discn=2$
and $\discn=3$ (see Fig. \ref{fig:energies}a). For clarity, we plot
$\Delta E=E(t,U)-E(0,U)$ divided by the number of spin orbitals.
The $\discn=3$ results are always lower in energy than the $\discn=2$
results, as is required by the variational principle. Furthermore,
the $\discn=3$ results resolve the well known deficiency of the $\discn=2$
results in the insulating regime. Interestingly, in the small and
large $U$ regimes, the quantity $\Delta E/(2\textrm{N}_{\textrm{orb}})$
is largely independent of $\textrm{N}_{\textrm{orb}}$, while in the
intermediate regime it decreases with $\textrm{N}_{\textrm{orb}}$.
Similar behavior is observed in the kinetic energy per orbital (see
Fig. \ref{fig:energies}b). The double occupancy $D$ determines the
interaction energy, and for convenience we plot the scaled double
occupancy $\tilde{D}=(D-D_{at})/(D_{0}-D_{at})$, where $D_{0}$ and
$D_{at}$ are the non-interacting and atomic double occupancy, respectively
(see Fig. \ref{fig:energies}c). In the small $U$ regime, the scaled
double occupancy decreases with $\textrm{N}_{\textrm{orb}}$, while
in the large $U$ regime it is independent of $\textrm{N}_{\textrm{orb}}$. 

\begin{figure}[tbph]
\includegraphics[width=1\columnwidth]{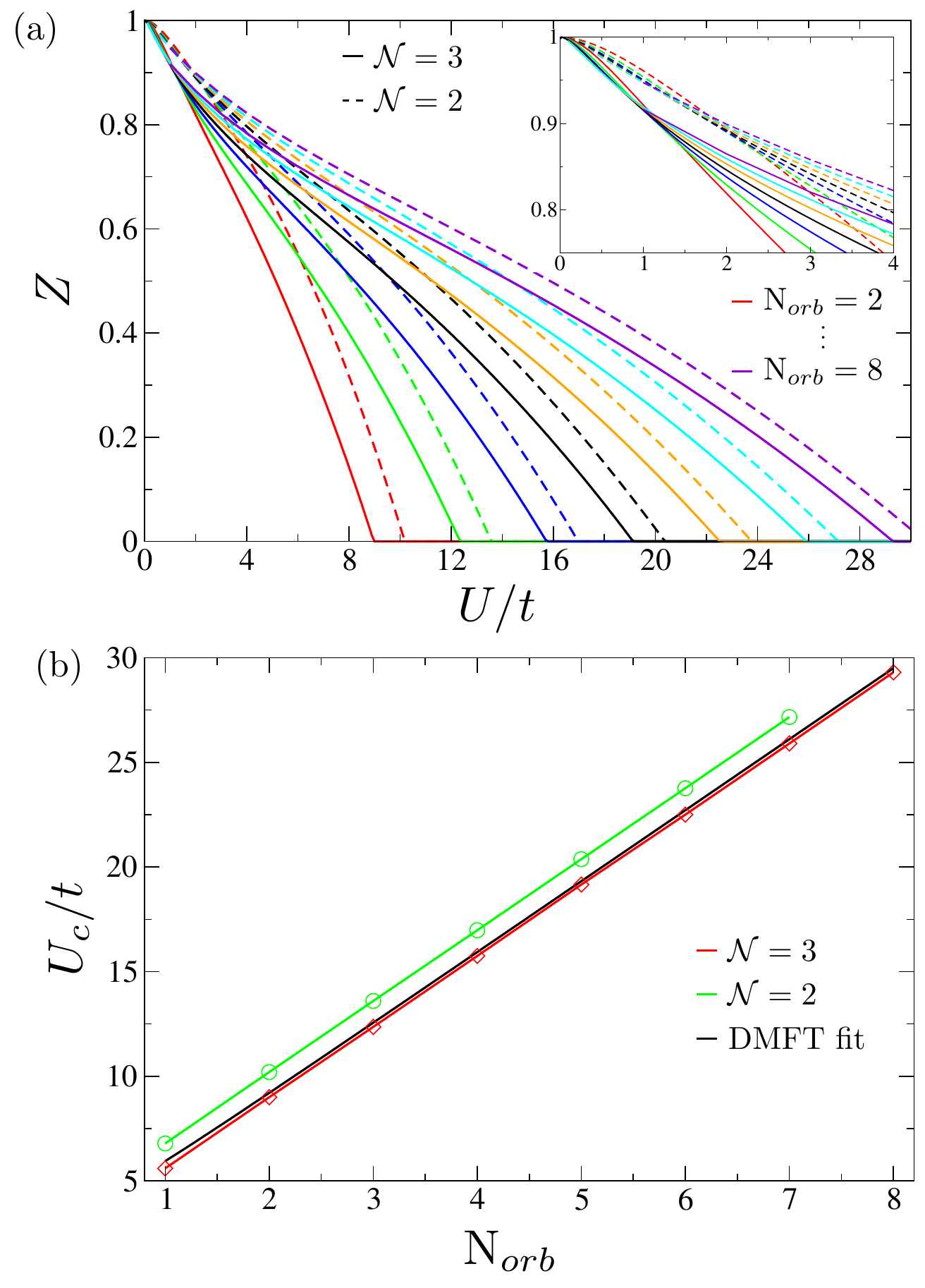}

\caption{Results for VDAT within the SCDA at $\discn=2$ and $\discn=3$ for
the SU(2N$_{\textrm{orb}}$) Hubbard model on the $d=\infty$ Bethe
lattice at half filling and $T=0$, with N$_{\textrm{orb}}=2-8$.
($a$) Quasiparticle weight. The legend is identical to Figure \ref{fig:Variational-Parameters}.
($b$) The critical value of $U$ for the MIT, denoted $U_{c}$, as
a function of $\textrm{N}_{\textrm{orb}}$. The DMFT curve is a plot
of a previously published fit to DMFT results \cite{Blumer2013085115},
given as $U_{c}/t=1.7(2\textrm{N}_{\textrm{orb}}+1)(1+0.166\textrm{N}_{\textrm{orb}}^{-1})$.
\label{fig:quasiparticle_weight} }
\end{figure}

It is also interesting to evaluate the quasiparticle weight as a function
of $U/t$, which determines the critical value of $U$ for the MIT
(see Fig. \ref{fig:quasiparticle_weight}a). The $\discn=2$ result
always produces a large quasiparticle weight than $\discn=3$, and
therefore produces a larger critical value of $U$ for the MIT. Interestingly,
for $\discn=3$ at approximately $U/t=1$, the quasiparticle weight
is insensitive to $\textrm{N}_{\textrm{orb}}$, and a similar effect
is observed for $\discn=2$ at a slightly large value of $U/t$. We
also examine the transition value $U_{c}/t$ as a function of $\textrm{N}_{\textrm{orb}}$
and compare to the previously published scaling relation \cite{Blumer2013085115}
$U_{c}/t=1.7(2\textrm{N}_{\textrm{orb}}+1)(1+0.166\textrm{N}_{\textrm{orb}}^{-1})$
extracted from DMFT calculations which use QMC as the impurity solver
(see Figure \ref{fig:quasiparticle_weight}$b$). The $\discn=2$
case recovers the previously published results of the Gutzwiller approximation
\cite{Lu19945687,Parcollet2002205102}, yielding a linear relation.
Interestingly, $\discn=3$ also produces a nearly linear result, but
the result is shifted downward, nearly coinciding with the DMFT extrapolation. 

\begin{figure}[tbph]
\includegraphics[width=1\columnwidth]{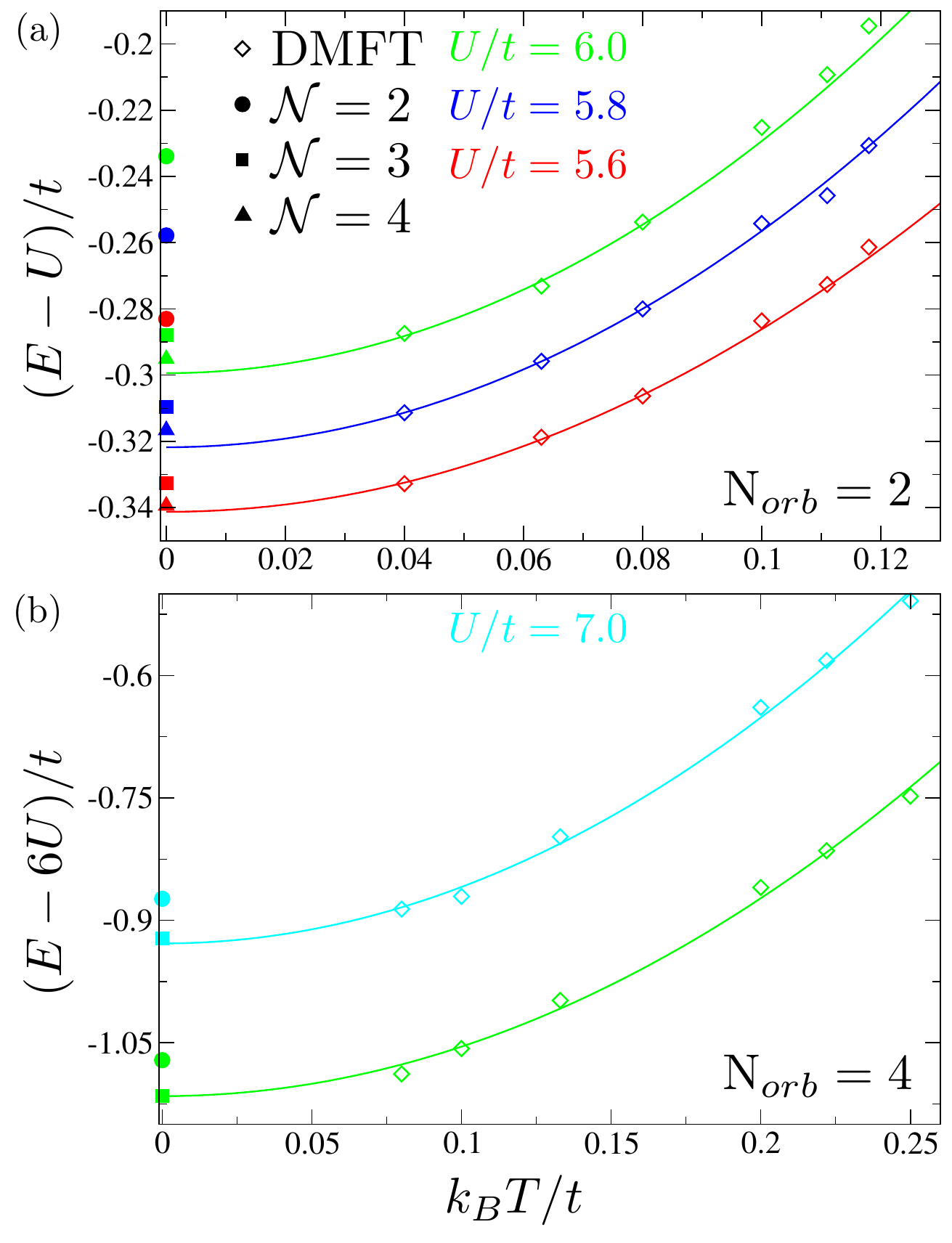}

\caption{The total energy of the SU(2N$_{\textrm{orb}}$) Hubbard model on
the $d=\infty$ Bethe lattice at half filling from published DMFT
calculations solved using QMC at a finite temperature \cite{Blumer2013085115}
and VDAT within the SCDA at zero temperature using $\discn=2-4$,
with N$_{\textrm{orb}}=2$ in panel $a$ and N$_{\textrm{orb}}=4$
in panel $b$. The lines are a fit to the DMFT results assuming $E(T)-E(0)\propto T^{2}$.\label{fig:qmc_half_fill}}
\end{figure}

We now proceed to compare the total energy at half filling and zero
temperature from VDAT and DMFT. Given that the previously published
DMFT results are at a finite temperature \cite{Blumer2013085115},
a quadratic fit assuming $E(T)-E(0)\propto T^{2}$ was used to extrapolate
to zero temperature (see Fig. \ref{fig:qmc_half_fill}). We were only
able to examine select cases where the QMC data sufficiently resembled
a quadratic. For $\textrm{N}_{\textrm{orb}}=2$, we present VDAT results
for $\discn=2-4$, which were previously published in Ref. \cite{Cheng2022205129},
while for $\textrm{N}_{\textrm{orb}}=4$ we present VDAT results for
$\discn=2-3$. As required by the variational principle, the energy
within VDAT monotonically decreases with increasing $\discn$. The
dramatic improvement of $\discn=3$ over $\discn=2$ is clearly illustrated,
and it should be recalled that these two cases have a similar computational
cost.

\begin{figure}[tbph]
\includegraphics[width=1\columnwidth]{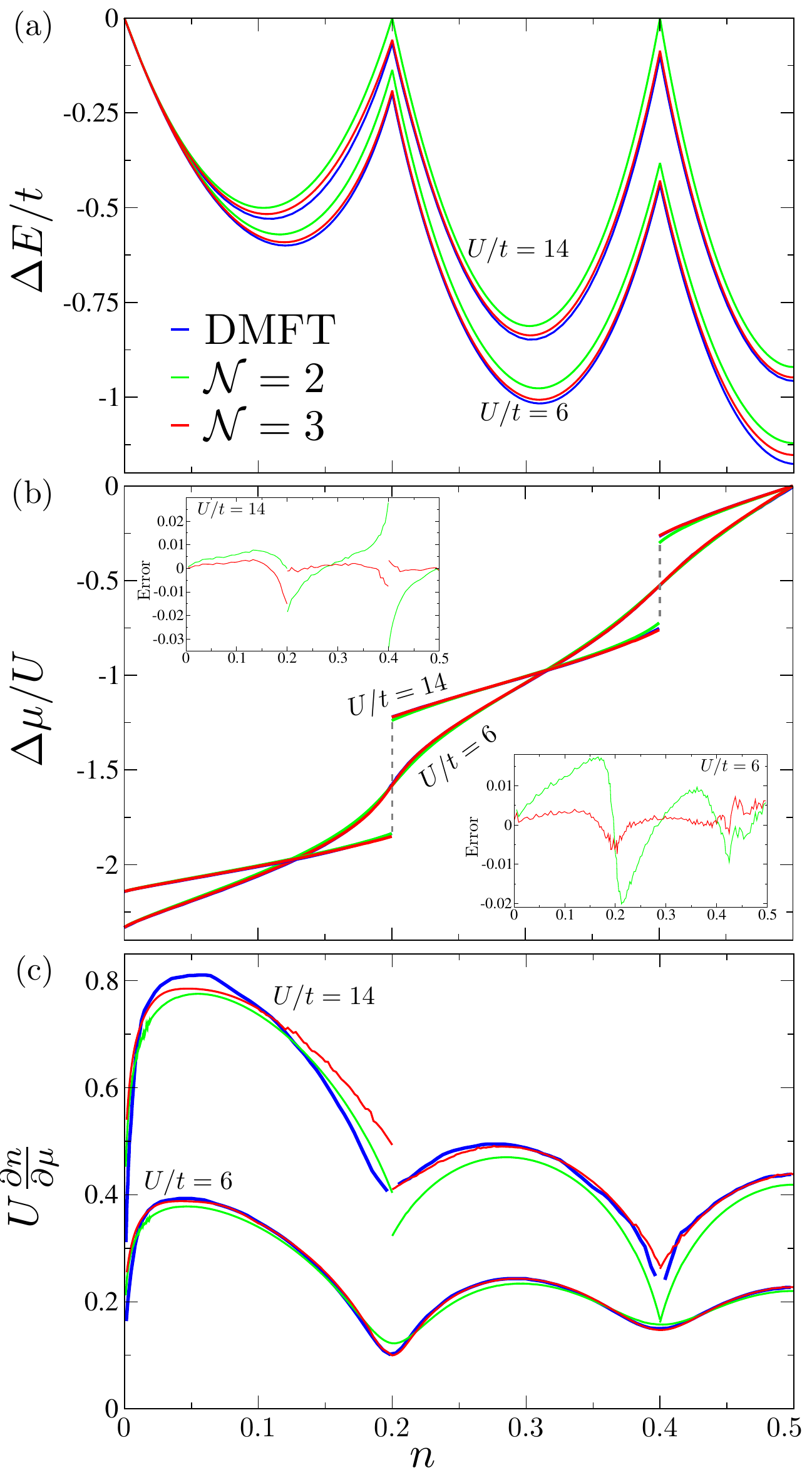}

\caption{Doping dependent results of the SU(2N$_{\textrm{orb}}$) Hubbard model
on the $d=\infty$ Bethe lattice at zero temperature from published
DMFT calculations solved using NRG \cite{Lee2018165143} and VDAT
within the SCDA using $\discn=2$ and $\discn=3$, with 2N$_{\textrm{orb}}=5$.
(panel $a$) Total energy difference $\Delta E$ vs. density, where
$\Delta E=E(t,U)-E(0,U)$. The DMFT curve is obtained by integrating
the chemical potential over density. (panel $b$) $\Delta\mu/U$ vs.
density, where $\Delta\mu=\mu-2U$. The insets plot the absolute error
in $\Delta\mu/U$ vs. density. (panel $c$) The derivative $\partial n/\partial\mu$
times $U$ vs. the density. \label{fig:nrg_compare}}
\end{figure}

Finally, we evaluate the doping dependence of the total energy, in
addition to the corresponding first and second derivatives, using
VDAT with $\discn=2$ and $\discn=3$. We compare to a previously
published DMFT study which used the NRG impurity solver \cite{Lee2018165143}
to study $2\textrm{N}_{\textrm{orb}}=5$ for $U/t=6$ and $U/t=14$
(see Fig. \ref{fig:nrg_compare}), where $U/t=6$ is a metal at all
densities and $U/t=14$ undergoes an MIT at the integer fillings of
$n=0.2$ and $n=0.4$. We first compare the total energy, where the
variational principle guarantees that the energy monotonically decreases
from $\discn=2$ to $\discn=3$ to the numerically exact solution
given by DMFT solved within NRG (see Fig. \ref{fig:nrg_compare},
panel $a$). It should be noted the DMFT study did not provide the
total energy, and we obtained it by numerically integrating the chemical
potential over the density. For clarity, we plot $\Delta E/t$ where
$\Delta E=E(t,U)-E(0,U)$ and $E(t,U)$ is the total energy evaluated
at a given density. Overall, $\discn=3$ yields a dramatic improvement
over $\discn=2$, especially at integer fillings. Interestingly, the
trends in the absolute error of the total energy are notably different
for $\discn=2$ and $\discn=3$, where the former has the largest
absolute error at integer filling while the latter has the largest
absolute error midway between integer fillings. The trend for $\discn=3$
might be attributed to the efficacy of the kinetic projector for capturing
superexchange at integer filling. We proceed to compare the chemical
potential, which is the derivative of the energy with respect to the
density, as a function of density (see Fig. \ref{fig:nrg_compare},
panel $b$). Within VDAT, the chemical potential was obtained using
finite difference to take the first derivative of the total energy
with respect to the density, and a grid spacing of 0.002 was used
for $0.02<n_{\alpha\sigma}\le0.5$ while 0.0001 was used for $n_{\alpha\sigma}<0.02$.
For clarity, we plot $\Delta\mu/U$ where $\Delta\mu=\mu-2U$. The
Mott transition can clearly be identified as a discontinuity in the
chemical potential at integer fillings. While $\discn=2$ is reasonable
overall, there are clear discrepancies near the MIT (see insets for
absolute error in $\Delta\mu/U$), and $\discn=3$ largely resolves
these issues. However, it should be recalled that unlike the total
energy, the convergence of the chemical potential is not necessarily
monotonic in $\discn$. For example, in the case of $U/t=14$ and
$n\rightarrow0.2^{-}$, the total energy for $\discn=2$ is substantially
larger than $\discn=3$, yet the chemical potential for $\discn=2$
is much closer to the exact solution. To further scrutinize these
same results from another viewpoint, we examine $U\partial n/\partial\mu$
as a function of density (see Fig. \ref{fig:nrg_compare}$c$). Within
VDAT, $\partial\mu/\partial n$ was obtained using finite difference
to take the second derivative of the total energy, and the same grid
spacing was used as in the case of the chemical potential. For the
low density region $n_{\alpha\sigma}<0.1$, the $\discn=3$ results
were smoothed using a spline interpolation. Similar to the chemical
potential, the convergence of this quantity is not necessarily monotonic
in $\discn$, and the same conclusions can be drawn. 

\section{Conclusions}

In this paper, we proposed a gauge constrained algorithm to evaluate
the SPD ansatz at $\discn=3$ within the SCDA for the multi-orbital
Hubbard model. The key feature of this algorithm is that it automatically
satisfies the self-consistency condition of the SCDA using the gauge
freedom of the SPD, greatly facilitating the minimization over the
variational parameters. Interestingly, the gauge constrained algorithm
yields a simple analytical form of the single particle density matrix
of the optimized SPD ansatz. The convenient mathematical form of the
gauge constrained algorithm greatly simplifies the implementation
of VDAT within the SCDA when treating a large number of orbitals.
In order to showcase the power of the gauge constrained algorithm,
we studied the SU(2N$_{\textrm{orb}}$) Hubbard model at zero temperature
on the Bethe lattice in $d=\infty$ for $\textrm{N}_{\textrm{orb}}=2-8$,
and compare to numerically exact DMFT results where available. While
the symmetry of the SU(2N$_{\textrm{orb}}$) Hubbard model greatly
reduces the computational cost for solving the DMFT impurity model,
computational limitations still restrict most studies to relatively
small values of N$_{\textrm{orb}}$. A DMFT study using a numerical
renormalization group impurity solver presented results up to $2\textrm{N}_{\textrm{orb}}=5$
\cite{Lee2018165143}, while a study using a quantum Monte-Carlo impurity
solver presented finite temperature results up to $\textrm{N}_{\textrm{orb}}=8$
\cite{Blumer2013085115}; though the latter results were at relatively
high temperatures and appear to have nontrivial stochastic error.
Therefore, our successful execution of $\textrm{N}_{\textrm{orb}}=8$
showcases the utility of the gauge constrained algorithm for executing
VDAT within the SCDA at $\discn=3$. At half filling, we evaluated
the kinetic energy, interaction energy, and quasiparticle weight as
a function of $U/t$. For the doped case, we evaluated the density
as a function of the chemical potential, in addition to the derivative,
at various values of $U/t$. As expected, $\discn=3$ yields a dramatic
improvement over $\discn=2$, at a similar computational cost. The
successful computation of the ground state energy for the $\textrm{N}_{\textrm{orb}}=7$
Hubbard model on a single processor core in under one hour demonstrates
the viability of VDAT to study realistic $f$-electron systems. The
technical developments in this work are a key step forward towards
studying realistic Hamiltonians of complex strongly correlated electron
materials. 

\section{Acknowledgements}

We thank Shuxiang Yang for useful discussions about the manuscript.
This work was supported by the Columbia Center for Computational Electrochemistry.
This research used resources of the National Energy Research Scientific
Computing Center, a DOE Office of Science User Facility supported
by the Office of Science of the U.S. Department of Energy under Contract
No. DE-AC02-05CH11231.

\appendix

\section{Solving multi-orbital Hubbard model with a density constraint\label{sec:app_density_constraint}}

It is often desirable to solve a Hamiltonian with fixed densities
for the spin orbitals, which can be efficiently executed by reparametrizing
the variational parameters $u$. We begin by realizing that the vector
space associated with $u$ can be constructed as a direct product
of two dimensional vector spaces associated with each spin orbital.
For an operator in the compound space $\barhat[O]=\prod_{\alpha\sigma}\barhat[O]_{\alpha\sigma}$,
the representation in the $u$ basis can be constructed as
\begin{equation}
(\barhat[O])_{u}=(\barhat[O]_{1\uparrow})_{u;1\uparrow}\otimes\dots\otimes(\barhat[O]_{N_{\textrm{orb}}\downarrow})_{u;N_{\textrm{orb}}\downarrow}.\label{eq:dp_operator_ubasis}
\end{equation}
Using this relation, equations \ref{eq:identity_in_compound} and
\ref{eq:gloc_representation} are recast as
\begin{align}
\left(\barhat[1]\right)_{u}= & \left(\barhat[1]\right)_{u;1\uparrow}\otimes\cdots\otimes\left(\barhat[1]\right)_{u;N_{\textrm{orb}}\downarrow},\label{eq:dp_identity}\\
(\barhat[a]_{\alpha\sigma}^{\dagger\left(\tau\right)}\barhat[a]_{\alpha\sigma}^{\left(\tau'\right)})_{u}= & \left(\barhat[1]\right)_{u;1\uparrow}\otimes\cdots\nonumber \\
 & \otimes(\barhat[a]_{\alpha\sigma}^{\dagger\left(\tau\right)}\barhat[a]_{\alpha\sigma}^{\left(\tau'\right)})_{u;\alpha\sigma}\dots\otimes\left(\barhat[1]\right)_{u;N_{\textrm{orb}}\downarrow},
\end{align}
where 
\begin{align}
 & [\left(\barhat[1]\right)_{u;\alpha\sigma}]_{\Gamma_{\alpha\sigma}\Gamma'_{\alpha\sigma}}=p_{\alpha\sigma}\left(\Gamma_{\alpha\sigma},\Gamma_{\alpha\sigma}'\right),\\
 & [(\barhat[a]_{\alpha\sigma}^{\dagger(\tau)}\barhat[a]_{\alpha\sigma}^{(\tau')})_{u;\alpha\sigma}]_{\Gamma_{\alpha\sigma}\Gamma'_{\alpha\sigma}}=g_{\alpha\sigma}^{\tau\tau'}\left(\Gamma_{\alpha\sigma},\Gamma'_{\alpha\sigma}\right),
\end{align}
where $p_{\alpha\sigma}\left(\Gamma_{\alpha\sigma},\Gamma_{\alpha\sigma}'\right)$
and $g_{\alpha\sigma}^{\tau\tau'}\left(\Gamma_{\alpha\sigma},\Gamma'_{\alpha\sigma}\right)$
are defined in equations \ref{eq:p_elements} and \ref{eq:g_elements},
respectively. The relevant matrices which will be needed to constrain
the orbital density are 

\begin{align}
 & \left(\barhat[1]\right)_{u;\alpha\sigma}=\left(\begin{array}{cc}
\mathcal{G}_{\alpha\sigma,1,2}^{2}+\frac{1}{4} & \frac{1}{4}-\mathcal{G}_{\alpha\sigma,1,2}^{2}\\
\frac{1}{4}-\mathcal{G}_{\alpha\sigma,1,2}^{2} & \mathcal{G}_{\alpha\sigma,1,2}^{2}+\frac{1}{4}
\end{array}\right),\\
 & \left(\barhat[a]_{\alpha\sigma}^{\dagger\left(1\right)}\barhat[a]_{\alpha\sigma}^{\left(1\right)}\right)_{u;\alpha\sigma}=\left(\begin{array}{cc}
0 & 0\\
\frac{1}{4}-\mathcal{G}_{\alpha\sigma,1,2}^{2} & \mathcal{G}_{\alpha\sigma,1,2}^{2}+\frac{1}{4}
\end{array}\right),\\
 & \left(\barhat[a]_{\alpha\sigma}^{\dagger\left(2\right)}\barhat[a]_{\alpha\sigma}^{\left(2\right)}\right)_{u;\alpha\sigma}=\left(\begin{array}{cc}
0 & \frac{1}{4}-\mathcal{G}_{\alpha\sigma,1,2}^{2}\\
0 & \mathcal{G}_{\alpha\sigma,1,2}^{2}+\frac{1}{4}
\end{array}\right).
\end{align}
In summary, Eq. \ref{eq:dp_operator_ubasis} provides a simple mathematical
structure to construct $(\barhat[O])_{u}$.

We now proceed to reparametrize the variational parameters $u$. In
general, one can introduce a linear transformation over the variational
parameters as $u=Vw$, and by requiring $w^{T}(\barhat[O])_{w}w=u^{T}(\barhat[O])_{u}u,$
a new matrix form is obtained as 
\begin{equation}
(\barhat[O])_{w}\equiv V^{T}(\barhat[O])_{u}V.
\end{equation}
In order to preserve the direct product structure of $(\barhat[O])_{w}$,
the transformation is constructed as $V=V_{1\uparrow}\otimes\cdots\otimes V_{\textrm{N}_{\textrm{orb}}\downarrow}$,
resulting in 
\begin{align}
 & (\barhat[O])_{w}=(\barhat[O]_{1\uparrow})_{w;1\uparrow}\otimes\dots\otimes(\barhat[O]_{\textrm{N}_{\textrm{orb}}\downarrow})_{w;\textrm{N}_{\textrm{orb}}\downarrow},\\
 & (\barhat[O]_{\alpha\sigma})_{w;\alpha\sigma}\equiv V_{\alpha\sigma}^{T}(\barhat[O]_{\alpha\sigma})_{u;\alpha\sigma}V_{\alpha\sigma}.
\end{align}
In order to ensure that $(\barhat[1])_{w;\alpha\sigma}$ is the identity
matrix and the symmetric part of $(\barhat[a]_{\alpha\sigma}^{\dagger\left(1\right)}\barhat[a]_{\alpha\sigma}^{\left(1\right)})_{w;\alpha\sigma}$
is diagonal, we have 
\begin{equation}
\boldsymbol{V}_{\alpha\sigma}=\frac{1}{\sqrt{2}}\left(\begin{array}{cc}
\frac{1}{2\mathcal{G}_{\alpha\sigma,1,2}}+1 & 1-\frac{1}{2\mathcal{G}_{\alpha\sigma,1,2}}\\
1-\frac{1}{2\mathcal{G}_{\alpha\sigma,1,2}} & \frac{1}{2\mathcal{G}_{\alpha\sigma,1,2}}+1
\end{array}\right),
\end{equation}
thus completely defining the reparametrization. One of the necessary
reparameterized matrix elements is 
\begin{align}
\boldsymbol{n}_{w;\alpha\sigma} & \equiv\frac{1}{2}\left((\barhat[a]_{\alpha\sigma}^{\dagger\left(1\right)}\barhat[a]_{\alpha\sigma}^{\left(1\right)})_{w;\alpha\sigma}+(\barhat[a]_{\alpha\sigma}^{\dagger\left(1\right)}\barhat[a]_{\alpha\sigma}^{\left(1\right)})_{w;\alpha\sigma}^{T}\right)\nonumber \\
 & =\left(\begin{array}{cc}
-\frac{\left(1-2\mathcal{G}_{\alpha\sigma,1,2}\right){}^{2}}{8\mathcal{G}_{\alpha\sigma,1,2}} & 0\\
0 & \frac{\left(2\mathcal{G}_{\alpha\sigma,1,2}+1\right){}^{2}}{8\mathcal{G}_{\alpha\sigma,1,2}}
\end{array}\right),
\end{align}
and the others are provided in Supplementary Material \cite{supplementary}.

We now proceed to constrain the density for each spin orbital, and
we begin by considering the case where the interacting projector is
a non-interacting density matrix, which can be written as 

\begin{equation}
w_{\Gamma;0}^{2}=\prod_{\alpha\alpha}\left(\left(1-n_{\alpha\sigma;eff}\right)\left(1-\Gamma_{\alpha\sigma}\right)+n_{\sigma\alpha;eff}\Gamma_{\alpha\sigma}\right),
\end{equation}
where $n_{\alpha\sigma;eff}$ can be determined from 
\begin{align}
\textrm{Tr}\left(\boldsymbol{n}_{w;\alpha\sigma}\left(\begin{array}{cc}
1-n_{\alpha\sigma;eff} & 0\\
0 & n_{\sigma\alpha;eff}
\end{array}\right)\right) & =n_{\alpha\sigma},
\end{align}
which can be solved as 
\begin{equation}
n_{\alpha\sigma,\text{eff}}=\frac{2\left(2n_{\alpha\sigma}-1\right)\mathcal{G}_{\alpha\sigma,1,2}}{4\mathcal{G}_{\alpha\sigma,1,2}^{2}+1}+\frac{1}{2}.\label{eq:neff}
\end{equation}
Subsequently, $2^{2\textrm{N}_{\textrm{orb}}}-\left(1+2\textrm{N}_{\textrm{orb}}\right)$
variational parameters $x_{\eta}$ can be introduced to describe the
deviations from $w_{\Gamma;0}^{2}$, which do not change the density
or the normalization. It is then useful to define a $2^{2\textrm{N}_{\textrm{orb}}}\times2^{2\textrm{N}_{\textrm{orb}}}$
matrix $V_{\Gamma\eta}$ as
\begin{align}
 & V_{\Gamma i(\{\alpha_{1}\sigma_{1},\dots,\alpha_{n}\sigma_{n}\})}=\prod_{j=1}^{n}(\Gamma_{\alpha_{j}\sigma_{j}}-\frac{1}{2}),
\end{align}
where $i(\{\alpha_{1}\sigma_{1},\dots,\alpha_{n}\sigma_{n}\})=1,\dots,2^{2\textrm{N}_{\textrm{orb}}}$
is a convention for indexing all subsets of $\{1\uparrow,1\downarrow,\dots,\textrm{N}_{\textrm{orb}}\uparrow,\textrm{N}_{\textrm{orb}}\downarrow\}$,
and $n=0,\dots,2\textrm{N}_{\textrm{orb}}$ denotes the number of
spin orbitals contained in a given subset. A convenient convention
for sorting the subsets is first sorting by increasing cardinality
and then by the binary interpretation of the subset. The subsets with
cardinality greater than one form $2^{2\textrm{N}_{\textrm{orb}}}-\left(1+2\textrm{N}_{\textrm{orb}}\right)$
orthogonal vectors that do not change the normalization or the orbital
occupation. A similar approach has been used to represent the Bernoulli
distribution \cite{Teugels1990256}. A general interacting projector
that is constrained to the given orbital occupation can be parameterized
as 
\begin{equation}
w_{\Gamma}^{2}=w_{\Gamma;0}^{2}+\sum_{\eta=2+2\textrm{N}_{\textrm{orb}}}^{2^{2\textrm{N}_{\textrm{orb}}}}V_{\Gamma\eta}x_{\eta},
\end{equation}
where $x_{\eta}$ are real numbers that are constrained by the condition
that $\omega_{\Gamma}^{2}\ge0$. For example, $\textrm{N}_{\textrm{orb}}=1$
results in one independent variational parameter $x$, yielding

\begin{align}
 & w_{1}^{2}=\left(1-n_{\uparrow;eff}\right)\left(1-n_{\downarrow eff}\right)+\frac{1}{4}x,\\
 & w_{2}^{2}=\left(1-n_{\uparrow;eff}\right)n_{\downarrow eff}-\frac{1}{4}x,\\
 & w_{3}^{2}=n_{\uparrow;eff}\left(1-n_{\downarrow eff}\right)-\frac{1}{4}x,\\
 & w_{4}^{2}=n_{\uparrow;eff}n_{\downarrow eff}+\frac{1}{4}x,
\end{align}
where $x\in[x_{\textrm{min}},x_{\textrm{max}}]$ and
\begin{align}
 & x_{\textrm{min}}=-4\min((1-n_{\uparrow;eff})(1-n_{\downarrow eff}),n_{\uparrow;eff}n_{\downarrow eff}),\\
 & x_{\textrm{max}}=4\min((1-n_{\uparrow;eff})n_{\downarrow eff},n_{\uparrow;eff}(1-n_{\downarrow eff})).
\end{align}

Finally, the ground state energy can be obtained as
\begin{align}
E=\min_{\mathcal{G}_{12},x,\bm{b}}\Big( & \int dk\epsilon_{k\alpha\sigma}n_{k\alpha\sigma}\left(\bm{a},\bm{b}\right)+E_{loc}\left(\mathcal{G}_{12},x,\bm{b}\right)\Big),
\end{align}
where $x=\{x_{\eta}\}$ and $\bm{a}$ is determined from $\left\{ n_{\alpha\sigma}\right\} $,
$\mathcal{G}_{12}$, $x$, and $\bm{b}$. 

\section{the gauge constrained algorithm using general local projectors \label{appendix:limitations}}

In this paper, we have assumed that the interacting projector can
be written as a linear combination of diagonal Hubbard operators in
the basis $\alpha\sigma$, and that $\boldsymbol{\mathcal{G}}$ is
diagonal in basis $\alpha\sigma$. Here we outline how to treat the
general case, starting with the first assumption. A general local
interacting projector can be an arbitrary linear combination of all
possible Hubbard operators, including off-diagonal Hubbard operators.
A general Hubbard operator can be constructed as 
\begin{align}
\hat{P}_{i\Gamma}=\prod_{\alpha\sigma} & \Big(\delta_{\Gamma_{\alpha\sigma},0}(1-\hat{n}_{i\alpha\sigma})+\delta_{\Gamma_{\alpha\sigma},1}\hat{n}_{i\alpha\sigma}\nonumber \\
 & +\delta_{\Gamma_{\alpha\sigma},2}\hat{a}_{i\alpha\sigma}^{\dagger}+\delta_{\Gamma_{\alpha\sigma},3}\hat{a}_{i\alpha\sigma}\Big),\\
=\prod_{\alpha\sigma} & \hat{P}_{i\alpha\sigma;\Gamma_{\alpha\sigma}},\label{eq:general_projector}
\end{align}
where $\Gamma-1=\left(\Gamma_{1\uparrow}\dots\Gamma_{\textrm{N}_{\textrm{orb}}\downarrow}\right)_{4}$
and $\Gamma=1,\dots,4^{2\textrm{N}_{\textrm{orb}}}$. The most general
interacting projector can be constructed as $\hat{P}_{i}(u)=\sum_{\Gamma}u_{\Gamma}\hat{P}_{i\Gamma}$.
However, given that we require $\hat{P}_{i}(u)$ to obey certain symmetries
and conservation relations, some $u_{\Gamma}$ may be zero. In order
to evaluate $\langle\barhat[P]_{i}^{\left(1\right)}\barhat[P]_{i}^{\left(2\right)}\barhat[O]\rangle_{\barhat[\rho]_{loc;i,0}}$,
we first consider 
\begin{equation}
\barhat[O]=\prod_{\alpha\sigma}\barhat[O]_{\alpha\sigma},\label{eq:oproduct}
\end{equation}
where $\barhat[O]_{\alpha\sigma}$ is a single product in terms of
$\barhat[a]_{i\alpha\sigma}^{\dagger\left(\tau\right)}$ and $\barhat[a]_{i\alpha\sigma}^{\left(\tau\right)}$,
yielding 
\begin{align}
\left\langle \barhat[P]_{i}^{\left(1\right)}\barhat[P]_{i}^{\left(2\right)}\barhat[O]\right\rangle _{\barhat[\rho]_{loc;i,0}} & =\sum_{\Gamma\Gamma'}u_{\Gamma}u_{\Gamma'}\left\langle \barhat[P]_{i\Gamma}^{\left(1\right)}\barhat[P]_{i\Gamma'}^{\left(2\right)}\barhat[O]\right\rangle _{\barhat[\rho]_{loc;i,0}},
\end{align}
where
\begin{align}
 & \left\langle \barhat[P]_{i\Gamma}^{\left(1\right)}\barhat[P]_{i\Gamma'}^{\left(2\right)}\barhat[O]\right\rangle _{\barhat[\rho]_{loc;i,0}}\nonumber \\
 & =\left\langle \prod_{\alpha\sigma}\barhat[P]_{i\alpha\sigma;\Gamma_{\alpha\sigma}}^{\left(1\right)}\prod_{\alpha\sigma}\barhat[P]_{i\alpha\sigma;\Gamma'_{\alpha\sigma}}^{\left(2\right)}\prod_{\alpha\sigma}\barhat[O]_{\alpha\sigma}\right\rangle _{\barhat[\rho]_{loc;i,0}}\label{eq:p1p2o}\\
 & =\theta\left(\barhat[O],\Gamma,\Gamma'\right)\prod_{\alpha\sigma}\left\langle \barhat[P]_{i\alpha\sigma;\Gamma_{\alpha\sigma}}^{\left(1\right)}\barhat[P]_{i\alpha\sigma;\Gamma'_{\alpha\sigma}}^{\left(2\right)}\barhat[O]_{\alpha\sigma}\right\rangle _{\barhat[\rho]_{loc;i,0}},\label{eq:theta_p1p20}
\end{align}
where $\theta(\barhat[O],\Gamma,\Gamma')=\pm1$ and is determined
by tracking the sign when ordering the expression from Eq. \ref{eq:p1p2o}
to Eq. \ref{eq:theta_p1p20}. For a general operator $\barhat[O]$,
one can always decompose it into a sum of operators which has the
form of Eq \ref{eq:oproduct} and apply the above formulas.

In order to treat a general $\boldsymbol{\mathcal{G}}$ and a general
operator $\barhat[O]$, one must straightforwardly apply Wick's theorem
to evaluate expectation values \cite{Cheng2021195138}, though the
resulting gauge constrained algorithm will be more complicated. For
example, simple closed form equations such as Eq. \ref{eq:dens_dist}
may not be obtained, requiring a numerical minimization to obtain
the density distribution.

\section{Understanding how the $\mathcal{N}=3$ gauge constrained algorithm
with a restricted kinetic projector reduces to $\mathcal{N}=2$ }

In Ref. \cite{Cheng2021195138}, we illustrated how the SCDA at $\mathcal{N}=2$
using the Gutzwiller gauge recovers the Gutzwiller approximation.
In the present work where we address $\mathcal{N}=3$, the gauge constrained
algorithm uses a different type of gauge. Therefore, it is interesting
to see how the $\mathcal{N}=3$ gauge constrained algorithm with a
restricted kinetic projector can recover the Gutzwiller approximation.
In particular, the restricted kinetic projector will force the density
distribution to be flat both above and below the fermi surface. We
begin by assuming 
\begin{equation}
\bm{\mathcal{G}}_{\alpha\sigma}=\left(\begin{array}{ccc}
\frac{1}{2} & \frac{1}{2} & \frac{1}{2}\\
-\frac{1}{2} & \frac{1}{2} & \frac{1}{2}\\
-\frac{1}{2} & -\frac{1}{2} & \frac{1}{2}
\end{array}\right),
\end{equation}
which is motivated by the Gutzwiller gauge. The canonical discrete
action of $\bm{\mathcal{G}}_{\alpha\sigma}$ corresponds to an SPD
\cite{Cheng2021195138}, which is the product of three identity operators,
and we can write the $A$ block for interacting Green's function as
\begin{equation}
\boldsymbol{g}_{loc,\alpha\sigma;A}=\left(\begin{array}{cc}
n_{\alpha\sigma} & a_{\alpha\sigma}r_{\alpha\sigma}\\
-a_{\alpha\sigma}r_{\alpha\sigma} & n_{\alpha\sigma}
\end{array}\right),
\end{equation}
where $a_{\alpha\sigma}=\sqrt{\left(1-n_{\alpha\sigma}\right)n_{\alpha\sigma}}$
and $r_{\alpha\sigma}$ denotes the renormalization for the off-diagonal
elements of the A-block compared to the reference interacting Green's
function 
\begin{equation}
\boldsymbol{g}_{loc,\alpha\sigma;A;ref}=\left(\begin{array}{cc}
n_{\alpha\sigma} & a_{\alpha\sigma}\\
-a_{\alpha\sigma} & n_{\alpha\sigma}
\end{array}\right),
\end{equation}
which corresponds to the case where $\hat{P}$ is non-interacting,
denoted as $\hat{P}_{0}$ . The $\hat{P}_{0}$ is chosen such that
\begin{equation}
\frac{\text{Tr}\left(\hat{P}_{0}^{2}\hat{n}_{\alpha\sigma}\right)}{\text{Tr}\left(\hat{P}_{0}^{2}\right)}=\frac{\text{Tr}\left(\hat{P}^{2}\hat{n}_{\alpha\sigma}\right)}{\text{Tr}\left(\hat{P}^{2}\right)}=n_{\alpha\sigma},
\end{equation}
and $r_{\alpha\sigma}$ is given as

\begin{equation}
r_{\alpha\sigma}=\frac{\text{Tr}\left(\hat{P}\hat{a}_{\alpha\sigma}^{\dagger}\hat{P}\hat{a}_{\alpha\sigma}\right)}{\text{Tr}\left(\hat{P}^{2}\right)}/\frac{\text{Tr}\left(\hat{P}_{0}\hat{a}_{\alpha\sigma}^{\dagger}\hat{P}_{0}\hat{a}_{\alpha\sigma}\right)}{\text{Tr}\left(\hat{P}_{0}^{2}\right)},
\end{equation}

The point of introducing the reference $\hat{P}_{0}$ is to allow
comparison with the Gutzwiller approximation. Given that the local
energy can be computed as 
\begin{equation}
E_{loc}=\frac{\text{Tr}\left(\hat{P}^{2}\hat{H}_{loc}\right)}{\text{Tr}\left(\hat{P}^{2}\right)},
\end{equation}
which is the same as in the Gutzwiller approximation, the remaining
task is to confirm that the kinetic energy recovers the Gutzwiller
approximation when restricting the density distribution to be flat,
and confirm that the self-consistency of the SCDA is maintained. 

We begin by computing the A-block of the local integer time self-energy
as 

\begin{align}
\boldsymbol{S}_{loc,\alpha\sigma;A} & =\left(\bm{\mathcal{G}}_{\alpha\sigma;A}^{-1}-\boldsymbol{1}\right)^{-1}\left(\boldsymbol{g}_{loc,\alpha\sigma;A}^{-1}-\boldsymbol{1}\right),
\end{align}
which yields the integer time self-energy as
\[
\boldsymbol{S}_{loc,\alpha\sigma}=\left(\begin{array}{ccc}
S_{\alpha\sigma;11} & S_{\alpha\sigma;12} & 0\\
-S_{\alpha\sigma;12} & S_{\alpha\sigma;11} & 0\\
0 & 0 & 1
\end{array}\right),
\]
where
\begin{align*}
S_{\alpha\sigma;11} & =\frac{a_{\alpha\sigma}r_{\alpha\sigma}}{n_{\alpha\sigma}\left(n_{\alpha\sigma}-\left(n_{\alpha\sigma}-1\right)r_{\alpha\sigma}^{2}\right)},\\
S_{\alpha\sigma;12} & =\frac{1}{n_{\alpha\sigma}-\left(n_{\alpha\sigma}-1\right)r_{\alpha\sigma}^{2}}-1.
\end{align*}
Assuming the kinetic projector is the identity, corresponding to $\lambda_{k\alpha\sigma,1}=1/2$,
we get a flat distribution in both the $<$ and $>$ region, yielding
\begin{align}
 & n_{k\alpha\sigma}|{}_{k\in<}=n_{\alpha\sigma}\left(1-r_{\alpha\sigma}^{2}\right)+r_{\alpha\sigma}^{2},\\
 & n_{k\alpha\sigma}|{}_{k\in>}=n_{\alpha\sigma}\left(1-r_{\alpha\sigma}^{2}\right).
\end{align}
One can verify that the integral of the density distribution yields
the corresponding local density as 
\begin{equation}
\int_{<}n_{k\alpha\sigma}dk+\int_{>}n_{k\alpha\sigma}dk=n_{\alpha\sigma},
\end{equation}
and the real space renormalization of the hopping parameter, $Z_{\alpha\sigma}$,
can be computed as
\begin{equation}
Z_{\alpha\sigma}=n_{k\alpha\sigma}|{}_{k\in<}-n_{k\alpha\sigma}|{}_{k\in>}=r_{\alpha\sigma}^{2},
\end{equation}
which recovers the Gutzwiller approximation.

We now confirm that the self-consistency condition of the SCDA is
maintained. Computing the local integer time Green's function yields 

\begin{equation}
g'_{loc}=\left(\begin{array}{ccc}
n_{\alpha\sigma} & a_{\alpha\sigma}r_{\alpha\sigma} & a_{\alpha\sigma}r_{\alpha\sigma}\\
-a_{\alpha\sigma}r_{\alpha\sigma} & n_{\alpha\sigma} & n_{\alpha\sigma}\\
-a_{\alpha\sigma}r_{\alpha\sigma} & n_{\alpha\sigma}-1 & n_{\alpha\sigma}
\end{array}\right),
\end{equation}
which yields 
\begin{equation}
\mathcal{G}'=\left(\begin{array}{ccc}
\frac{1}{2} & \frac{1}{2} & \frac{1}{2}\\
-\frac{1}{2} & \frac{1}{2} & \frac{1}{2}\\
-\frac{1}{2} & -\frac{1}{2} & \frac{1}{2}
\end{array}\right),
\end{equation}
which indicates that the SCDA self-consistency is satisfied. 

%\bibliographystyle{plain}
%\bibliography{main,software}

\end{document}